\documentclass[useAMS,usenatbib]{mn2e}
\usepackage{graphicx}
\usepackage{appendix}
\usepackage{lscape}
\usepackage{amsmath}  
\usepackage{subfig}     
\usepackage{amssymb}  
\usepackage{longtable}

\def\lsim{\,\lower2truept\hbox{${<\atop\hbox{\raise4truept\hbox{$\sim$}}}$}\,}
\def\gsim{\,\lower2truept\hbox{${> \atop\hbox{\raise4truept\hbox{$\sim$}}}$}\,}

\title[{\it H}$-$ATLAS: magnifications and sizes of lensed galaxies]
{The {\it Herschel}$-$ATLAS: magnifications and physical sizes
of $500\,\micron$-selected strongly lensed galaxies
}
\author[Enia A. et al.]{A. Enia$^{1}$\thanks{E-mail: andreafrancescomaria.enia@studenti.unipd.it},
M. Negrello$^{2}$,
M. Gurwell$^{3}$,
S. Dye$^{4}$,
G. Rodighiero$^{1}$,
M. Massardi$^{5}$,
\newauthor
G. De Zotti$^{6}$,
A. Franceschini$^{1}$,
A. Cooray$^{7}$,
P. van der Werf$^{8}$,
M. Birkinshaw$^{9}$,
\newauthor
M.~J. Micha{\l}owski$^{10}$,
I. Oteo$^{11,12}$
\vspace{4mm}\\
$^{1}$Dipartimento di Fisica e Astronomia, Universit{\`a} di Padova, vicolo dell’Osservatorio 3, I-35122 Padova, Italy\\
$^{2}$School of Physics and Astronomy, Cardiff University, The Parade, Cardiff, CF24 3AA, UK\\
$^{3}$Harvard-Smithsonian Center for Astrophysics, MA 02138 Cambridge, USA\\
$^{4}$School of Physics and Astronomy, University of Nottingham, University Park, Nottingham NG7 2RD, UK \\
$^{5}$INAF, Istituto di Radioastronomia, Via Gobetti 101, I-40129 Bologna, Italy\\
$^{6}$INAF - Osservatorio Astronomico di Padova, Vicolo dell'Osservatorio 5, I-35122, Padova, Italy \\
$^{7}$Department of Physics and Astronomy, University of California, CA 92697 Irvine, USA \\
$^{8}$Leiden Observatory, Leiden University, P.O. Box 9513, NL-2300, RA Leiden, the Netherlands \\
$^{9}$HH Willis Physics Laboratory, University of Bristol, Tyndall Avenue, Bristol BS8 1TL, UK \\
$^{10}$Astronomical Observatory Institute, Faculty of Physics, Adam Mickiewicz University, ul.~S{\l}oneczna 36, 60-286 Pozna{\'n}, Poland \\
$^{11}$Institute for Astronomy, University of Edinburgh, Royal Observatory, Blackford Hill, Edinburgh EH9 3HJ UK \\
$^{12}$European Southern Observatory, Karl-Schwarzschild-Str. 2, 85748 Garching, Germany \\
}

\pagerange{\pageref{firstpage}--\pageref{lastpage}} \pubyear{2018}
\date{Accepted 2017 December 21. Received 2017 December 21; in original form 2017 November 10}

\begin{document}

\label{firstpage}

\maketitle

\begin{abstract}
We perform lens modelling and source reconstruction of Submillimeter Array (SMA) data for a sample of 12 strongly lensed galaxies selected at 500$\micron$
in the {\it Herschel} Astrophysical Terahertz Large Area Survey ({\it H}-ATLAS). A previous analysis of the same dataset used a single S{\'e}rsic profile
to model the light distribution of each background galaxy. Here we model the source brightness distribution with an adaptive pixel scale scheme,
extended to work in the Fourier visibility space of interferometry. We also present new SMA observations for seven other candidate lensed galaxies
from the {\it H}-ATLAS sample. Our derived lens model parameters are in general consistent with previous findings. However, our estimated magnification
factors, ranging from 3 to 10, are lower. The discrepancies are observed in particular where the reconstructed source hints at the presence of multiple
knots of emission. We define an effective radius of the reconstructed sources based on the area in the source plane where emission is detected above
5$\sigma$. We also fit the reconstructed source surface brightness with an elliptical Gaussian model. We derive a median value $r_{\rm eff}\sim1.77\,$kpc and a
median Gaussian full width at half maximum $\sim1.47\,$kpc. After correction for magnification, our sources have intrinsic star formation rates
SFR$\,\sim900-3500\,{\rm M}_{\sun}{\rm yr}^{-1}$, resulting in a median star formation rate surface density $\Sigma_{\rm SFR}\sim132\,{\rm M}_{\sun}{\rm yr}^{-1}\,$kpc$^{-2}$ (or
$\sim218\,{\rm M}_{\sun}{\rm yr}^{-1}\,$kpc$^{-2}$ for the Gaussian fit). This is consistent with what observed for other star forming galaxies at similar redshifts, and is
significantly below the Eddington limit for a radiation pressure regulated starburst.
\end{abstract}

\begin{keywords}
gravitational lensing: strong -- instrumentation: interferometers -- galaxies: structure
\end{keywords}

\section{Introduction}
%
The samples of strongly lensed galaxies generated by wide-area extragalactic surveys performed at sub-millimetre (sub-mm) to millimetre (mm) 
wavelengths \citep{2010Sci...330..800N,2017MNRAS.465.3558N,2013ApJ...762...59W,2013Natur.495..344V,2015A&A...582A..30P,2016ApJ...823...17N}, 
with the {\it Herschel} space observatory \citep{2010A&A...518L...1P}, the South Pole Telescope \citep{2011PASP..123..568C} and the {\it Planck}
satellite \citep{2015A&A...581A.105C} provide a unique opportunity to study and understand the physical properties of the most violently star 
forming galaxies at redshifts $z > 1$. In fact, the magnification induced by gravitational lensing makes these objects extremely bright and, 
therefore, excellent targets for spectroscopic follow-up observations aimed at probing the physical conditions of the interstellar medium in the 
distant Universe \citep[e.g.][]{2011MNRAS.415.3473V,2012ApJ...757..135L,2012ApJ...752..152H,2011A&A...530L...3O,2013A&A...551A.115O,
2017arXiv170105901O, 2016A&A...595A..80Y}. At the same time, the increase in the angular sizes of the background sources due to lensing allows us to 
explore the structure and dynamics of distant galaxies down to sub-kpc scales 
\citep[e.g.][]{2010Natur.464..733S, 2015ApJ...806L..17S, 2015MNRAS.451L..40R, 2015MNRAS.452.2258D}.

In order to be able to fully exploit these advantages, it is crucial to reliably reconstruct the background galaxy from the observed lensed 
images. The process of source reconstruction usually implies 2an analytic assumption about the surface brightness of the source, for example by 
adopting S{\'e}rsic or Gaussian profiles
\citep[e.g][]{2008ApJ...682..964B, 2013ApJ...779...25B, 2015ApJ...812...43B, 2014ApJ...797..138C, 2016ApJ...826..112S}.
However this approach can be risky, particularly for objects with often complex, clumpy, morphologies like those exhibited by sub-mm/mm selected
dusty star forming galaxies (DSFG) when observed at resolutions of tens of milliarcseconds
\citep[e.g.][]{2010Natur.464..733S, 2011ApJ...742...11S, 2015MNRAS.452.2258D}.

Sophisticated lens modelling and source reconstruction techniques have recently been developed to overcome this problem. 
\cite{1996ApJ...465...64W} introduced the idea of a pixellated background source, where each pixel value is treated as an independent parameter,
thus avoiding any assumption on the shape of the source surface brightness distribution. \cite{2003ApJ...590..673W} showed that with this approach
the problem of reconstructing the background source, for a fixed lens mass model, is reduced to the inversion of a matrix. The best-fitting lens
model parameters can then be explored via standard Monte Carlo techniques. In order to avoid unphysical solutions, the method introduces a
regularization term that forces a certain degree of smoothness in the reconstructed source. The weight assigned to this regularization term is set
by Bayesian analysis \citep{2006MNRAS.371..983S}. Further improvements to the method include pixel sizes adapting to the lens magnification pattern
\citep{2005ApJ...623...31D, 2009MNRAS.392..945V,2015MNRAS.452.2940N} and non-smooth lens mass models \citep{2009MNRAS.392..945V,2016ApJ...823...37H}
in order to detect dark matter sub-structures in the foreground galaxy acting as the lens.

The method has been extensively implemented in the modelling of numerous lensed galaxies observed with instruments such as the {\it Hubble} Space
Telescope and the {\it Keck} telescope
\citep[e.g.][]{2004ApJ...611..739T, 2006ApJ...649..599K, 2010MNRAS.408.1969V, 2008MNRAS.388..384D, 2014MNRAS.440.2013D, 2015MNRAS.452.2258D}.
For DSFGs, high resolution imaging data usable for lens modelling can mainly be achieved by interferometers at the sub-mm/mm wavelengths where these
sources are bright. Since the lensing  galaxy is usually a massive elliptical, there is virtually no contamination from the lens at those wavelengths.
However, an interferometer does not directly measure the surface brightness of the source, but instead it samples its Fourier transform,
named the {\it visibility function}. As such, the lens modelling of interferometric images needs to be carried out in Fourier space in order to
minimize the effect of correlated noise in the image domain and to properly account for the undersampling of the signal in Fourier space, which
produces un-physical features in the reconstructed image.

Here, we start from the adaptive source pixel scale method of \cite{2015MNRAS.452.2940N} and extend it to work directly in the Fourier space to
model the Sub-Millimeter Array (SMA) observations of a sample of 12 lensed galaxies discovered in the {\it Herschel} Astrophysical Terahertz Large
Area Survey \citep[{\it H}-ATLAS][]{2010PASP..122..499E}; eleven of these sources were previously modelled by 
\citet[][B13 hereafter]{2013ApJ...779...25B} assuming a S{\'e}rsic profile for the light distribution of the background galaxy. We reassess their 
findings with our new approach and also present SMA follow-up observations of 7 more candidate lensed galaxies from the 
{\it H}-ATLAS \citep{2017MNRAS.465.3558N}, although we attempted lens modelling for only one of them, where multiple images can be resolved in the
data.

The paper is organized as follows: Section\,\ref{sec:data} presents the sample and the SMA observations.
Section\,\ref{sec:methodology} describes the methodology used for the lens modeling and its application to interferometric data. In
Section\,\ref{sec:results} we present and discuss our findings, with respect to the results of B13 and other results from the literature. 
Conclusions are summarized in Section\,\ref{sec:conclusions}. Throughout the paper we adopt the Planck13 cosmology \citep{2014A&A...571A..16P},
with H$_0 = 67$ km s$^{-1}$ Mpc$^{-1}$, $\Omega_m = 0.32$, $\Omega_{\Lambda} = 0.68$, and assume a \cite{2001MNRAS.322..231K} initial mass function.

\begin{table*}
\centering
\caption{
  List of {\it H}-ATLAS lensed galaxies with SMA imaging data selected for the lens modelling and source reconstruction. Most are taken 
  from Bussmann et al. (2013), excluding group/cluster scale lenses and sources which are not clearly resolved into multiple images by the SMA.
  The list also includes candidate lensed galaxies from N17 for which we have obtained new SMA observations. However only one of them is clearly 
  resolved into multiple images because of the limited resolution achieved and therefore only this object, HATLAS J120127.6-014043, is considered
  for the lens modelling. Reading from left to right, columns following the identifier are: redshifts of the lens and of the background galaxy
  (from N17; when no spectroscopic redshift is available the photometric one is provided instead, in italic style), SPIRE/{\it Herschel} flux
  densities at 250, 350 and 500$\micron$ (from N17), flux density from the SMA, array configuration of the observations performed with the SMA
  (SUB$=$sub-compact, COM$=$compact, EXT$=$extended, VEX$=$very extended).
  }
\label{tab:sample}
\tiny
\scriptsize
\begin{tabular}{llllllll}
\hline
{\it H}-ATLAS IAU Name  & $z_{\rm opt}$   & $z_{\rm sub-mm}$  & $F_{250}$   & $F_{350}$   & $F_{500}$   & $F_{\rm SMA}$   & SMA Array   \\
                    &           &       & (mJy)     & (mJy)     & (mJy)     & (mJy)     & Configuration  \\
\hline
\multicolumn{6}{l}{\bf SMA data from Bussmann et al. (2013)} \\
HATLASJ083051.0+013225  & 0.6261+1.0002 & 3.634 & 248.5$\pm$7.5 & 305.3$\pm$8.1 & 269.1$\pm$8.7 &  76.6$\pm$2.0 & COM+EXT   \\   
HATLASJ085358.9+015537  & -     & 2.0925  & 396.4$\pm$7.6 & 367.9$\pm$8.2 & 228.2$\pm$8.9 &  50.6$\pm$2.6 & COM+EXT+VEX \\   
HATLASJ090740.0$-$004200 & 0.6129 & 1.577   & 477.6$\pm$7.3 & 327.9$\pm$8.2 & 170.6$\pm$8.5 &  20.3$\pm$1.8 & COM+EXT   \\   
HATLASJ091043.0$-$000322 & 0.793    & 1.786   & 420.8$\pm$6.5 & 370.5$\pm$7.4 & 221.4$\pm$7.8 &  24.4$\pm$1.8 & COM+EXT+VEX \\   
HATLASJ125135.3+261457  & -       & 3.675   & 157.9$\pm$7.5 & 202.3$\pm$8.2 & 206.8$\pm$8.5 &  64.5$\pm$3.4 & COM+EXT   \\   
HATLASJ125632.4+233627  & 0.2551  & 3.565   & 209.3$\pm$7.3 & 288.5$\pm$8.2 & 264.0$\pm$8.5 &  85.5$\pm$5.6 & COM+EXT   \\   
HATLASJ132630.1+334410  & 0.7856    & 2.951   & 190.6$\pm$7.3 & 281.4$\pm$8.2 & 278.5$\pm$9.0 &  48.3$\pm$2.1 & EXT     \\   
HATLASJ133008.4+245900  & 0.4276  & 3.1112  & 271.2$\pm$7.2 & 278.2$\pm$8.1 & 203.5$\pm$8.5 &  49.5$\pm$3.4 & COM+EXT   \\   
HATLASJ133649.9+291800  & -     & 2.2024  & 294.1$\pm$6.7 & 286.0$\pm$7.6 & 194.1$\pm$8.2 &  37.6$\pm$6.6 & SUB+EXT+VEX \\   
HATLASJ134429.4+303034  & 0.6721  & 2.3010  & 462.0$\pm$7.4 & 465.7$\pm$8.6 & 343.3$\pm$8.7 &  55.4$\pm$2.9 & COM+EXT+VEX \\   
HATLASJ142413.9+022303  & 0.595     & 4.243   & 112.2$\pm$7.3 & 182.2$\pm$8.2 & 193.3$\pm$8.5 & 101.6$\pm$7.4 & COM+EXT+VEX \\   
\hline
\multicolumn{6}{l}{\bf New SMA observations}  \\ 
HATLASJ120127.6$-$014043  & -                  & {\it 3.80$\pm$0.58} &  67.4$\pm$6.5  & 112.1$\pm$7.4  & 103.9$\pm$7.7  &  52.4$\pm$3.2 & COM+EXT \\
HATLASJ120319.1$-$011253  & -                & {\it 2.70$\pm$0.44} & 114.3$\pm$7.4  & 142.8$\pm$8.2  & 110.2$\pm$8.6  &  40.4$\pm$2.4 & COM+EXT \\ 
HATLASJ121301.5$-$004922  & {\it 0.191$\pm$0.080}& {\it 2.35$\pm$0.40} & 136.6$\pm$6.6  & 142.6$\pm$7.4  & 110.9$\pm$7.7  &  23.4$\pm$1.7 & COM+EXT \\
HATLASJ132504.3+311534    & {\it 0.58$\pm$0.11}  & {\it 2.03$\pm$0.36} & 240.7$\pm$7.2  & 226.7$\pm$8.2  & 164.9$\pm$8.8  &  35.2$\pm$2.2 & COM   \\  
HATLASJ133038.2+255128    & {\it 0.20$\pm$0.15}  & {\it 1.82$\pm$0.34} & 175.8$\pm$7.4  & 160.3$\pm$8.3  & 104.2$\pm$8.8  &  19.1$\pm$1.9 & COM   \\
HATLASJ133846.5+255054    & {\it 0.42$\pm$0.10}  & {\it 2.49$\pm$0.42} & 159.0$\pm$7.4  & 183.1$\pm$8.2  & 137.6$\pm$9.0  &  27.4$\pm$2.5 & COM   \\
HATLASJ134158.5+292833    & {\it 0.217$\pm$0.015}& {\it 1.95$\pm$0.35} & 174.4$\pm$6.7  & 172.3$\pm$7.7  & 109.2$\pm$8.1  &  20.9$\pm$1.5 & COM   \\
\hline
\end{tabular}
\end{table*}

\section{Sample and SMA data}\label{sec:data}

\subsection{Sample selection}

Our starting point is the sample of candidate lensed galaxies presented in \citet[][N17 hereafter]{2017MNRAS.465.3558N} which comprises 80 objects
with $F_{500}\ge100\,$mJy extracted from the full {\it H}-ATLAS survey. We kept only the sources in that sample with available SMA 870$\,\mu$m 
continuum follow-up observations, which are presented in B13 \citep[but see also][]{2010Sci...330..800N,2012ApJ...756..134B}. There are 21 in total.
We excluded three cluster scale lenses for which the lens modelling is complicated by the need for three or more mass 
models for the foreground objects (HATLASJ114637.9$-$001132, HATLASJ141351.9$-$000026, HATLASJ132427.0+284449). We also removed those sources
where the multiple images are not fully resolved by the SMA and therefore are not usable for source reconstruction, i.e. HATLASJ090302.9$-$014127, 
HATLASJ091304.9$-$005344, HATLASJ091840.8+023048, HATLASJ113526.2$-$014606, HATLASJ144556.1$-$004853, HATLASJ132859.2+292326. 
Finally we have not considered in our analysis HATLASJ090311.6+003907, also known as SDP.81, which has been extensively modelled
using high resolution data from the Atacama Large Millimetre Array
\citep{2015ApJ...808L...4A,2015MNRAS.451L..40R,2015PASJ...67...93H,2015ApJ...806L..17S,2015PASJ...67...72T,2015MNRAS.452.2258D,2016ApJ...823...37H}.
We have added an extra source to our sample, HATLASJ120127.6$-$014043, for which we recently obtained new SMA data (see Sec.
\ref{subsec:SMA_data}). Therefore, our final sample comprises 12 objects, which are included in Table \ref{tab:sample}.

\begin{figure*}
\begin{centering}
\hspace{-0.5cm}
\includegraphics[width=18cm]{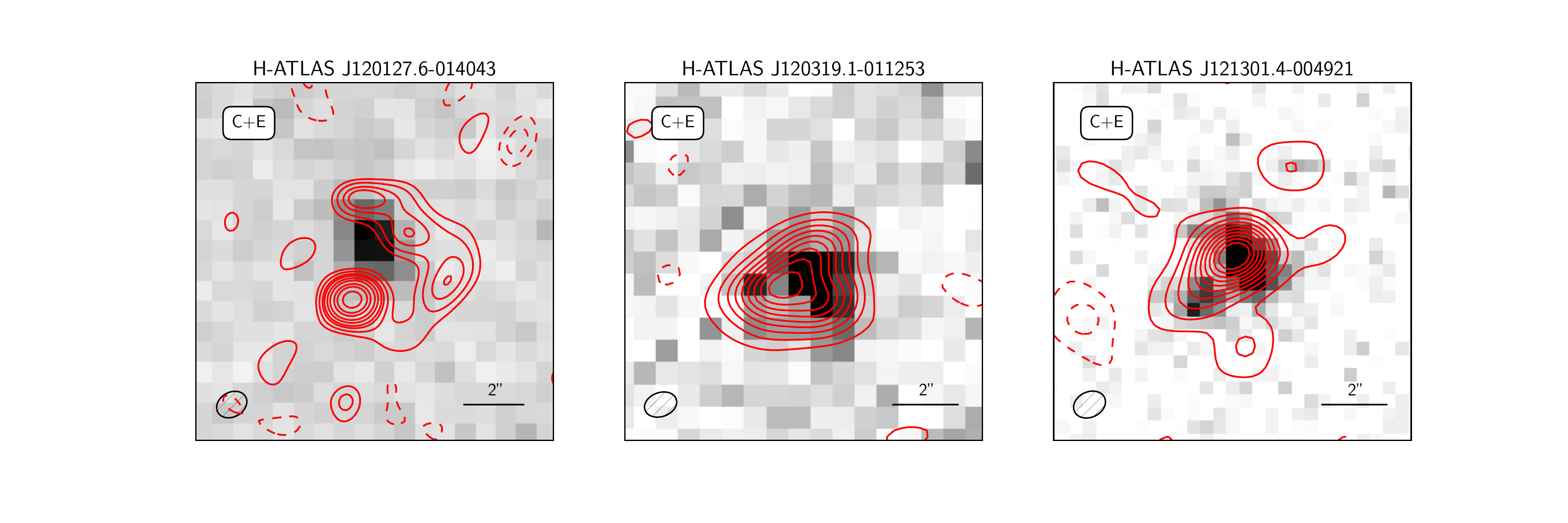} 
\end{centering} \\
\begin{centering}
\vspace{-0.8cm}
\includegraphics[width=18cm]{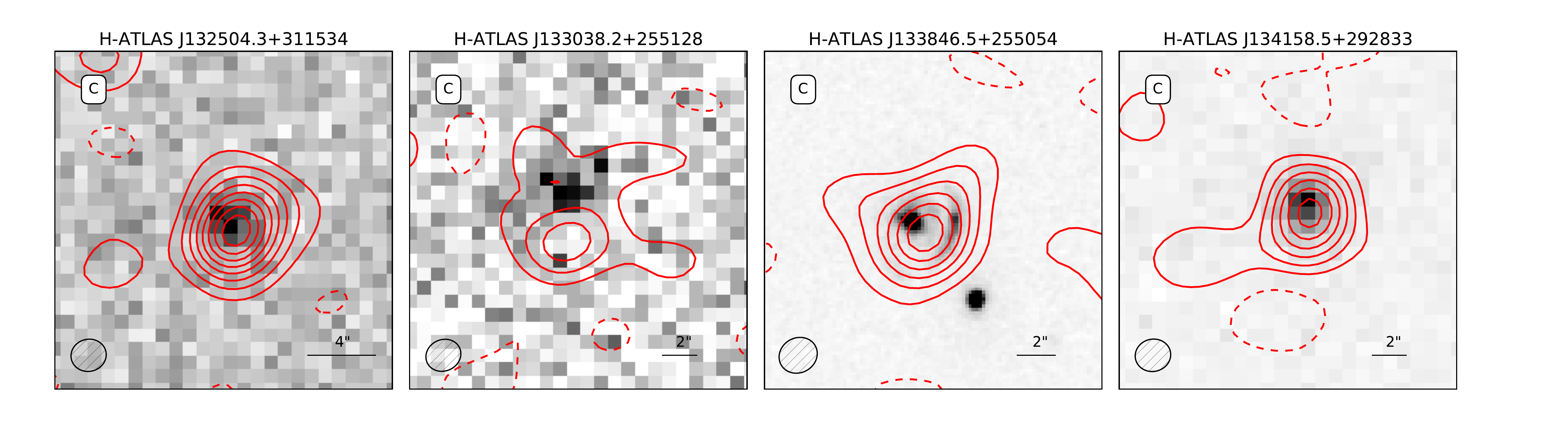}
\end{centering}
\vspace{-0.7cm}
\caption{
  New SMA 870\,$\micron$ follow-up observations (red contours, starting at $\pm2\sigma$ and increasing by factors of two) of seven {\it H}-ATLAS 
  candidate lensed galaxies from the N17 sample. The three sources in the top panels were observed in both compact and extended array 
  configurations, while the four sources in the bottom panels only have data obtained in compact configuration. The SMA’s synthesized beam is
  shown in the lower left corner of each panel. The background images, in grey-scale, show the best available optical/near-IR data and come from 
  the Kilo Degree Survey
  \citep[KiDS;][{\it r} band at $0.62\,\mu$m for HATLASJ120127.6$-$014043 and HATLASJ121301.4$-$004921]{2015A&A...582A..62D}, the VISTA 
  Kilo-Degree Infrared Galaxy Survey \citep[VIKING;][{\it K}$_{s}$ band at 2.2\,$\micron$ for HATLAS120319.1$-$011253]{2013Msngr.154...32E},
  the UK Infrared Deep Sky Survey Large Area Survey \citep[UKIDSS-LAS;][{\it K}$_s$ band at $2.2\,\mu$m for HATLASJ132504.3+311534 and 
  HATLASJ133038.2+255128; {\it Y} band at $1.03\,\mu$m for HATLASJ134158.5+292833]{2007MNRAS.379.1599L} and the {\it HST} Wide Field Camera 3
  (WFC3) at $1.6\,\mu$m (for HATLASJ133846.5+255054). 
    }
\label{fig:SMA_newsources}
\end{figure*}

\subsection{SMA data}\label{subsec:SMA_data}

The SMA data used here have been presented in B13 [but see also \cite{2010Sci...330..800N}]. They were obtained as part of a large proposal carried
out over several semesters using different array configurations from compact (COM) to very-extended (VEX), reaching a spatial resolution of
$\sim0.5^{\prime\prime}$, with a typical integration time of one to two hours on-source, per configuration. We refer the reader to B13 for details
concerning the data reduction. \\

Between December 2016 and March 2017 we carried out new SMA continuum observations at 870$\,\mu$m of a further seven candidate lensed galaxies from
the N17 sample (proposal ID: 2016B-S003  PI: Negrello). These targets were selected for having a reliable optical/near-IR counterpart with colours and
redshift inconsistent with those derived from the {\it Herschel}/SPIRE photometry. Therefore they are very likely to be lensing events, where the
lens is clearly detected in the optical/near-IR. They are listed in Table\,\ref{tab:sample}, and shown in Fig.\,\ref{fig:SMA_newsources}. Since we
were awarded B grade tracks, not all of the observations were executed. Thus while all seven sources were observed in COM configuration, only for three
did we also obtain data in the extended (EXT) configuration. 

Observations of HATLAS120127.6$-$014043, HATLAS120319.1$-$011253, and HATLAS121301.5$-$004922 were obtained in COM configuration (maximum baselines
$\sim$77m) on 29 December 2016. The weather was very good and stable, with a mean atmospheric opacity $\tau_{\rm 225GHz}=0.06$ (translating to 1\,
mm precipitable water vapor). All eight antennas participated, with 6\,GHz of continuum bandwidth per sideband in each of two polarizations (for 
the equivalent of 24\,GHz total continuum bandwidth). The central frequency of the observations was 344 GHz (870$\,\mu$m). The target observations 
were interleaved over a roughly 8 hour transit period, resulting in 100 to 110\,minutes of on-source integration time for each target (with the 
balance spent on bandpass and gain calibration using the bright, nearby radio source 3C273). The absolute flux scale was determined using 
observations of Callisto. Imaging the visibility data with a natural weighting scheme produced a synthesized beam with full-width at half maximum
(FWHM) $\sim2^{\prime\prime}$, and all three targets were detected with high confidence, with achieved image RMS values of 1.3\,mJy\,beam$^{-1}$.
Data for these three sources were combined with later higher resolution data (see below). 

Observations of HATLASJ132504.3+311534, HATLASJ133038.2+255128, HATLASJ133846.5+255054, HATLASJ134158.5+292833 were obtained in the COM 
configuration on 02 January 2017. The weather was good and stable, with $\tau_{\rm 225GHz}=0.07$ (corresponding to 1.2 mm precipitable water 
vapor). All eight antennas participated, with 6\,GHz of continuum bandwidth per sideband in each of two polarizations (for the equivalent of 24\,
GHz total continuum bandwidth). The mean frequency of the observations was 344\,GHz (870$\,\mu$m). The target observations were interleaved over a
roughly four hour rising to transit period, resulting in 40\,minutes of on-source integration time for three targets (HATLASJ134158.5+292833 only 
received 30\,min). Gain calibration was performed using observations of the nearby radio source 3C286, while bandpass and absolute flux scale were 
determined using observations of Callisto. Imaging the visibility data produced a synthesized beam with ${\rm FWHM}\sim2^{\prime\prime}$, and all 
four targets were detected with high confidence, with image RMS values of 1.5-1.7\,mJy\,beam$^{-1}$. 

HATLAS120127.6$-$014043, HATLAS120319.1$-$011253, and HATLAS121301.5$-$004922 were also observed in the EXT configuration
(maximum baselines ~220 m) on 29 March 2017. The weather was excellent and fairly stable, with a mean $\tau_{\rm 225GHz}$ of 0.04 rising to 0.05
(translating to 0.65-0.8 mm precipitable water vapor). All eight antennas participated, now with 8 GHz of continuum bandwidth per sideband in each
of two polarizations (for the equivalent of 32 GHz total continuum bandwidth); however, on one antenna only one of the two receivers was 
operational, resulting in a small loss of signal-to-noise ($\sim3\%$). The mean frequency of the observations was 344\,GHz (870$\,\mu$m). The
target observations were interleaved over a roughly six hour mostly rising transit period, resulting in 75 to 84 minutes of on-source integration 
time for each target. Bandpass and phase calibration observations were of 3C273, and the absolute flux scale was determined using observations of 
Ganymede. These extended configuration data were then imaged jointly with the compact configuration data from 29 December 2016. For each source, 
the synthesized resolution is roughly $1.0^{\prime\prime}\times 0.8^{\prime\prime}$, and the rms in the combined data maps ranges from 800
to 900 $\mu$Jy\,beam$^{-1}$. 

There is evidence of extended structure in several of our new targets, even from the COM data alone (e.g. HATLASJ133038.2+255128 and 
HATLASJ133846.5+255054); however only in HATLASJ120127.6$-$014043, which benefits from EXT data, are the typical multiple images of a lensing events 
clearly detected and resolved.\\
The measured 870\,$\micron$ flux density for each source is reported in Table\,\ref{tab:sample}. It was computed by adding up the signal inside a 
customized aperture that encompasses the source emission. The quoted uncertainties correspond to the root-mean-square variation of the primary-beam
corrected signal measured within the same aperture in 100 random positions inside the region defined by the primary beam of the instrument.\\
As explained in Section\,\ref{sec:methodology}, our lens modelling and source reconstruction are performed on the SMA data by adopting a natural
weighting scheme. The SMA {\it dirty} images obtained with this scheme are shown in the left panels of Fig.\,\ref{fig:lens_modelling_results}.

\section{Lens modelling and source reconstruction}\label{sec:methodology}

In order perform the lens modeling and to reconstruct the intrinsic morphology of the background galaxy, we follow the Regularized Semilinear
Inversion (SLI) method introduced by \cite{2003ApJ...590..673W}, which assumes a pixelated source brightness distribution. It also introduces a 
regularization term to control the level of smoothness of the reconstructed source. The method was improved by \cite{2006MNRAS.371..983S} using 
Bayesian analysis to determine the optimal weight of the regularization term and by \cite{2015MNRAS.452.2940N} with the introduction of a source 
pixelization that adapts to the lens model magnification. Here we adopt all these improvements and extend the method to deal with 
interferometric data. \\
We provide below a summary of the SLI method, but we refer the reader to \cite{2003ApJ...590..673W} for more details. 

\subsection{The adaptive semilinear inversion method}

The image plane (IP) and the source plane (SP), i.e. the planes orthogonal to the line-of-sight of the observer to the lens containing the lensed
images and the background source, respectively, are gridded into pixels whose values represent the surface brightness counts. In the IP, the pixel
values are described by an array of elements $d_j$, with $j=1,...,J$, and associated statistical uncertainty $\sigma_j$, while in the SP the unknown
surface brightness counts are represented by the array of elements $s_i$, with $i=1,...,I$. For a fixed lens mass model, the image plane is mapped to
the source plane by a unique rectangular matrix $f_{ij}$. The matrix contains information on the lensing potential, via the deflection angles, and on
the smearing of the images due to convolution with a given point spread function (PSF). In practice, the element $f_{ij}$ corresponds to the surface
brightness of the $j$-th pixel in the lensed and PSF-convolved image of source pixel $i$ held at unit surface brightness. The vector, $\textbf{S}$, of
elements $s_{i}$ that best reproduces the observed IP is found by minimizing the merit function
\begin{equation}
\label{eq:chisq_image}
G = \frac{1}{2} \chi^2 = \frac{1}{2}\sum_{j=1}^{J} \left( \dfrac{\sum_{i=1}^{I} s_i f_{ij} - d_j}{\sigma_j} \right) ^2.
\end{equation}
It is easy to show that the solution to the problem satisfies the matrix equation 
\begin{equation}
  \textbf{F} \cdot \textbf{S} = \textbf{D},
\end{equation}
where $\textbf{D}$ is the array of elements $D_{i}=\sum_{j=1}^{J} (f_{ij} d_j)/\sigma_j^2$ and $\textbf{F}$ is a symmetric matrix of elements
$F_{ik}=\sum_{j=1}^{J} (f_{ij} f_{kj})/ \sigma_j^2$. Therefore, the most likely solution for the source surface brightness counts can be obtained 
via a matrix inversion
\begin{equation}
\label{eq:Snotregol}
  \textbf{S} = \textbf{F}^{-1}\textbf{D}.
\end{equation}
However, in this form, the method may produce unphysical results. In fact, each pixel in the SP behaves independently from the others and, 
therefore, the reconstructed source brightness profile may show severe discontinuities and pixel-to-pixel variations due to the noise in the image 
being modelled. In order to overcome this problem a {\it prior} on the parameters $s_{i}$ is assumed, in the form of a regularization term,
$E_{\rm reg}$, which is added to the merit function in Eq.\,(\ref{eq:chisq_image}). This forces a smooth variation in the value of nearby pixels in
the SP:
\begin{equation}
\label{eq:chisq_image_reg}
  G_{\lambda} = \frac{1}{2}\chi^2 + \lambda E_{\rm reg} = \frac{1}{2}\chi^2 + \lambda \frac{1}{2}\textbf{S}^{T}\textbf{H}\textbf{S},
\end{equation}
where $\lambda$ is a {\it regularization constant}, which controls the strength of the regularization, and $\textbf{H}$ is the {\it regularization
matrix}. We have chosen a form for the regularization term $E_{\rm reg}$ that preserves the matrix formalism [see Eq.\,(\ref{eq:Ereg})]. The minimum of
the merit function in Eq.\,(\ref{eq:chisq_image_reg}) satisfies the condition
\begin{equation}
  [\textbf{F} + \lambda\textbf{H}]\cdot \textbf{S} = \textbf{D},
\end{equation}
and, therefore, can still be derived via a matrix inversion
\begin{equation}
\label{eq:SPsolution}
  \textbf{S} = [\textbf{F}+\lambda \textbf{H}]^{-1} \textbf{D}.
\end{equation}
The presence of the regularization term ensures the existence of a physical solution for any sensible regularization scheme.\\
The value of the regularization constant is found by maximizing the Bayesian evidence\footnote{Assuming a a flat prior on
$\log{\lambda}$.} $\epsilon$ \citep{2006MNRAS.371..983S}
\begin{eqnarray}
\label{eq:bayesianevidence}
2 \ln{[\epsilon(\lambda)]} & = & - G_{\lambda}(\textbf{S})  -  \ln [\det(\textbf{F} + \lambda \textbf{H})]  \nonumber \\
& & + \ln [\det (\lambda \textbf{H})] - \sum_{j=1}^{J} \ln{(2\pi \sigma_j^2)},
\end{eqnarray}
$\textbf{S}$ representing here the set of $s_{i}$ values obtained from Eq.\,(\ref{eq:SPsolution}) for a given $\lambda$. \\
The errors on the reconstructed source surface brightness distribution, for a fixed mass model, are given by the diagonal terms of the covariance 
matrix \citep{2003ApJ...590..673W}:
\begin{equation}
\label{eq:evidence}
\sigma_{ik}^2 = \sum_{j=1}^J \sigma_j^2 \dfrac{\partial s_i}{\partial d_j} \dfrac{\partial s_k}{\partial d_j} = 
R_{ik} - \lambda \sum_{l=1}^{I} R_{il}[\textbf{R}\textbf{H}]_{kl},
\end{equation}
where $\textbf{R}=[\textbf{F} + \lambda\textbf{H}]^{-1}$. We use this expression to draw signal-to-noise ratio contours in the reconstructed SP in
Fig.\,\ref{fig:lens_modelling_results} for the best fit lens model. 

Eqs\,(\ref{eq:SPsolution})-(\ref{eq:evidence}) allow us to derive the SP solution for a fixed lens mass model. However the parameters that best 
describe the mass distribution of the lens are also to be determined. This is achieved by exploring the lens parameter space and computing each
time the evidence in Eq.\,(\ref{eq:evidence}) marginalized over $\lambda$, i.e. $\epsilon=\int \epsilon(\lambda)P(\lambda)d\lambda$, where
$P(\lambda)$ is the probability distribution of the values of the regularization constant for a given lens model. The best-fitting values of the lens 
model parameters are those that maximize $\epsilon$. We follow \cite{2006MNRAS.371..983S} by approximating $P(\lambda)$ with a delta 
function centered around the value $\tilde{\lambda}$ that maximizes Eq.\,(\ref{eq:evidence}), so that $\epsilon\simeq\epsilon(\tilde{\lambda})$. \\

The pixels in the SP that are closer to the lens caustics are multiply imaged over different regions in the IP, and  therefore benefit from better
constraints during the source reconstruction process, compared to pixels located further away from the same lines. 
As a consequence, the noise in the reconstructed source surface brightness distribution significantly varies across the SP. At the same time, 
in highly magnified regions of the SP the information on the source properties at sub-pixel scales is not fully exploited. In order to overcome
this issue we follow the adaptive SP pixelization scheme proposed by \cite{2015MNRAS.452.2940N}. For a fixed mass model the IP pixel centres are 
traced back to the SP and a k-means clustering algorithm\footnote{This is slightly different than the h-means clustering scheme adopted by
\cite{2015MNRAS.452.2940N}, though the same adopted in \cite{2017arXiv170505413D}.} is used to group them and to define new pixel centres in the SP.
These centres are then used to generate Voronoi cells, mainly for visualization purposes. Within this adaptive pixelization scheme we use a
gradient regularization term defined as:
\begin{equation}\label{eq:Ereg}
E_{\rm reg} = \sum_{i=1}^{I} \sum_{k=1}^{N_{v}(i)} (s_{i} - s_{k})^{2}
\end{equation}
where $N_{v}(i)$ are the counts members of the set of Voronoi cells that share at least one vertex with the $i$-th pixel.

\subsection{Modeling in the uv plane}

We extend the adaptive SLI formalism to deal with images of lensed galaxies produced by interferometers.\\
An interferometer correlates the signals of an astrophysical source collected by an array of antennas to produce a {\it visibility function}
$V(u,v)$, that is the Fourier transform of the source surface brightness $I(x,y)$ sampled at a number of locations in the Fourier space, or
$uv$-plane:
\begin{equation}
V(u,v) = \int\int A(x,y) I(x,y) e^{-2\pi i(ux+vy)}dxdy
\end{equation}
where $A$ is the effective collecting area of each antenna, i.e. the primary beam.\\
Because of the incomplete sampling of the $uv$-plane the image of the astrophysical source obtained by Fourier transforming the visibility function
will be affected by artifacts, such as side-lobes, and by correlated noise. Therefore, a proper source reconstruction performed on interferometric data
should be carried out directly in the $uv$-plane.\\
We define the merit function using the visibility function\footnote{Besides the presence of the regularization term, 
this definition of the merit function is exactly as in \cite{2013ApJ...779...25B}.}
\begin{eqnarray}
\label{eq:chisq_uv_reg}
G_{\lambda} & = & \frac{1}{2}\sum_{u,v}^{N_{\rm vis}} \left \vert \frac{V_{\rm model}(u,v) - V_{\rm obs}(u,v)}{\sigma(u,v)} \right \vert^{2} + \lambda \frac{1}{2}\textbf{S}^{T}\textbf{H}\textbf{S} \nonumber \\
& = & \frac{1}{2}\sum_{u,v}^{N_{\rm vis}} \left( \frac{V_{\rm model}^{\mathbb{R}}(u,v) - V_{\rm obs}^{\mathbb{R}}(u,v)}{\sigma(u,v)} \right)^{2} \nonumber \\
&  & + \frac{1}{2}\sum_{u,v}^{N_{\rm vis}}  \left( \frac{V_{\rm model}^{\mathbb{I}}(u,v) - V_{\rm obs}^{\mathbb{I}} (u,v)}{\sigma(u,v)} \right)^{2} \nonumber \\
& &  +  \lambda \frac{1}{2}\textbf{S}^{T}\textbf{H}\textbf{S},
\end{eqnarray}
where $N_{\rm vis}$ is the number of observed visibilities $V_{\rm obs}=V_{\rm obs}^{\mathbb{R}} + iV_{\rm obs}^{\mathbb{I}}$, while
$\sigma^{2}(u,v) = \sigma_{\rm real}(u,v)^{2} + \sigma_{\rm imag}^{2}(u,v)$, with $\sigma_{\rm real}$ and $\sigma_{\rm imag}$ representing the
1$\sigma$ uncertainty on the real and imaginary parts of $V_{\rm obs}$, respectively. With this definition of the merit function 
we are assuming a natural weighting scheme for the visibilities in our lens modelling. \\
Following the formalism of Eq.\,(\ref{eq:chisq_image}), we can introduce a rectangular matrix of {\it complex} elements
$\hat{f}_{jk}=\hat{f}^{\mathbb{R}}_{jk} + i\hat{f}^{\mathbb{I}}_{jk}$, with $k=1,..,N_{\rm vis}$ and $j=1,..,N$, $N$ being the number of pixels
in the SP. The term $\hat{f}_{jk}$ provides the Fourier transform of a source pixel of unit surface brightness at the $j$-th pixel position
and zero elsewhere, calculated at the location of the $k$-th visibility point in the $uv$-plane. The effect of the primary beam is also accounted
for in calculating $\hat{f}_{jk}$. Therefore, Eq.\,(\ref{eq:chisq_uv_reg}) can be re-written as
\begin{eqnarray}
\label{eq:chisq_uv_reg_rewritten}
G_{\lambda} & = & \frac{1}{2}\sum_{u,v}^{N_{\rm vis}} \left \vert \frac{ V_{\rm model}(u,v) - V_{\rm obs}(u,v) }{\sigma(u,v)} \right \vert^{2} + \lambda \frac{1}{2}\textbf{S}^{T}\textbf{H}\textbf{S} \nonumber \\
& = & \frac{1}{2}\sum_{k}^{N_{\rm vis}} \left( \frac{\sum_{j=1}^{N} s_j\hat{f}^{\mathbb{R}}_{jk} - V_{{\rm obs},k}^{\mathbb{R}}}{\sigma_{k}} \right)^{2} \nonumber \\
&  & + \frac{1}{2}\sum_{k}^{N_{\rm vis}}  \left( \frac{\sum_{j=1}^{N} s_j\hat{f}^{\mathbb{I}}_{jk} - V_{{\rm obs},k}^{\mathbb{I}}}{\sigma_{k}} \right)^{2} \nonumber \\
& &  +  \lambda \frac{1}{2}\textbf{S}^{T}\textbf{H}\textbf{S}.
\end{eqnarray}
In deriving this expression we have assumed that $\bf{S}$ is an array of real values, as it describes a surface brightness.\\
The set of $s_{i}$ values that best reproduces the observed IP can then be derived as in Eq\,(\ref{eq:SPsolution}):
\begin{equation}
\label{eq:SPsolution_uv}
  \textbf{S} = [\hat{\textbf{F}}+\lambda \textbf{H}]^{-1} \hat{\textbf{D}}.
\end{equation}
with the new matrices $\hat{\textbf{F}}$ and $\hat{\textbf{D}}$ defined as follows
\begin{eqnarray}
\hat{\textbf{F}}_{jk} = \sum_{l=1}^{N_{\rm vis}} \frac{\hat{f}^{\mathbb{R}}_{jl} \hat{f}^{\mathbb{R}}_{lk} +
\hat{f}^{\mathbb{I}}_{jl} \hat{f}^{\mathbb{I}}_{lk}}{\sigma_{l}^{2}} \\
\hat{\textbf{D}}_{j} = \sum_{l=1}^{N_{\rm vis}} \frac{\hat{f}^{\mathbb{R}}_{jl} V_{{\rm obs},l}^{\mathbb{R}} + 
\hat{f}^{\mathbb{I}}_{jl} V_{{\rm obs},l}^{\mathbb{I}}}{\sigma_{l}^{2}} 
\end{eqnarray}
The computation of the regularization constant is exactly as in Eq.\,(\ref{eq:bayesianevidence}) with $G_{\lambda}$, $\textbf{F}$ and $\sigma$
replaced by the corresponding quantities defined in this section.
\begin{table*}
\centering
\caption{
  Results of the modelling for the lens mass distribution, for which a SIE profile is assumed. The parameters of the model are: the normalization
  of the profile, expressed in terms of the Einstein radius ($\theta_{\rm E}$); the rotation angle ($\theta_{\rm L}$; measured counter-clockwise
  from West); the minor-to-major axis ratio ($q_{\rm L}$); the position of the lens centroid from the centre of the images in
  Fig.\,\ref{fig:lens_modelling_results}; the shear strength ($\gamma$) and the shear angle ($\theta_{\gamma}$; counter-clockwise from West).
}
\label{table:parameters}
\begin{tabular}{lccccccc}
\hline
IAUname         & $\theta_{\rm E}$& $\theta_{\rm L}$ & $q_{\rm L}$  &$\Delta x_{\rm L}$ & $\Delta y_{\rm L}$ & $\gamma$ & $\theta_{\gamma}$ \\
              & (arcsec)        & ($^{\circ}$)    &         & (arcsec)    & (arcsec)       &             &  ($^{\circ}$)  \\
\hline
HATLASJ083051.0+013225  & 0.31$\pm$0.03 &  38.5$\pm$ 7.5 & 0.33$\pm$0.07 & $-$0.49$\pm$0.04 & +0.07$\pm$ 0.04   &  -                     &  -   \\
            & 0.58$\pm$0.05 & 172.6$\pm$16.8 & 0.82$\pm$0.08 & +0.18$\pm$0.03   & $-$0.63$\pm$0.05  &  -                     &  -   \\
HATLASJ085358.9+015537  & 0.54$\pm$0.01 &  62.3$\pm$30.0 & 0.95$\pm$0.05 & $-$0.22$\pm$0.03 & +0.03$\pm$0.03    &  -                     &  -   \\
HATLASJ090740.0$-$004200& 0.65$\pm$0.02 & 143.7$\pm$ 7.0 & 0.75$\pm$0.07 & $-$0.09$\pm$0.02 & $-$0.06$\pm$0.05  &  -                     &  -   \\
HATLASJ091043.1$-$000321& 0.91$\pm$0.03 & 112.9$\pm$10.2 & 0.62$\pm$0.09 & 0.00$\pm$0.07  & +0.33$\pm$0.05  & 0.20$\pm$0.05 & 76.0$\pm$12.0  \\
HATLASJ120127.6$-$014043& 0.82$\pm$0.04 & 169.0$\pm$6.7 & 0.58$\pm$0.09 & +0.06$\pm$0.06    & +2.00$\pm$0.05  &  -                     &  -   \\
HATLASJ125135.4+261457  & 1.10$\pm$0.02 &  28.0$\pm$2.5 & 0.51$\pm$0.06 & $-$0.23$\pm$0.05  & +0.39$\pm$0.04  &  -                     &  -   \\
HATLASJ125632.7+233625  & 0.69$\pm$0.03 &  24.6$\pm$7.4 & 0.54$\pm$0.09 & $-$0.05$\pm$0.10  & $-$0.10$\pm$0.06  &  -                     &  -    \\
HATLASJ132630.1+334410  & 1.76$\pm$0.05 & 149.4$\pm$9.0 & 0.62$\pm$0.08 & $-$0.49$\pm$0.10  & +0.67$\pm$0.10  &  -                     &  -    \\
HATLASJ133008.4+245900  & 1.03$\pm$0.02 & 172.1$\pm$2.2 & 0.51$\pm$0.03 & $-$1.54$\pm$0.08  & +0.95$\pm$0.04  &  -                     &  -    \\
HATLASJ133649.9+291801  & 0.41$\pm$0.02 &  38.5$\pm$4.3 & 0.53$\pm$0.12 & +0.22$\pm$0.05    & +0.20$\pm$0.04  &  -                     &  -    \\
HATLASJ134429.4+303036  & 0.96$\pm$0.01 &  82.7$\pm$1.5 & 0.53$\pm$0.07 & +0.34$\pm$0.06  & +0.02$\pm$0.03  &  -                     &  -    \\
HATLASJ142413.9+022303  & 0.98$\pm$0.02 &  91.0$\pm$4.9 & 0.79$\pm$0.04 & +1.09$\pm$0.03  & +0.33$\pm$0.04  &  -                     &  -  \\
\hline
\end{tabular}
\end{table*}

\subsection{Lens model}

In order to compare our findings with the results presented in B13, we model the mass distribution of the lenses as a Singular Isothermal Ellipsoid
\citep[SIE;][]{1994A&A...284..285K}, i.e. we assume a density profile of the form $\rho\propto r^{-2}$, $r$ being the elliptical radius.
Our choice of a SIE over a more generic power-law profile, $\rho\propto r^{-\alpha}$, is also motivated by the results of the modelling of other
lensing systems from literature \cite[e.g.][]{2009MNRAS.399...21B,2014MNRAS.440.2013D,2015MNRAS.452.2258D,2017arXiv170505413D}, which show that
$\alpha\sim2$, and by the need of keeping to a minimum the number of free parameters. In fact, the resolution of the SMA data analyzed here is a
factor $\times3-4$ worse than the one provided by the optical and near-infrared imaging data $-$ mainly from the Hubble space telescope $-$ used in
the aforementioned literature.\\
However it is important to point out that a degeneracy between different lens model profiles can lead to  biased estimates of the source size and
magnification. In fact, as first discussed by \cite{1985ApJ...289L...1F}, a particular rescaling of the density profile of the lens, together with an isotropic scaling
of the source plane coordinates, produces exactly the same observed image positions and flux ratios (but different time delays). This is known as the
mass-sheet transformation (MST) and represents a special case of the more general source-position transformation described by \cite{2014A&A...564A.103S}.
\cite{2013A&A...559A..37S} showed that the MST is formally broken by assuming a power-law model for the mass distribution of the lens, although there is
no physical reason why the true lens profile should have such an analytic form. Furthermore, the power-law model is also affected by the $\sigma-q-\alpha$
degeneracy between the lens mass (expressed in terms of the 1D velocity dispersion $\sigma$), the axis ratio ($q$) and the slope ($\alpha$). In fact, as discussed 
in \cite{2015MNRAS.452.2940N}, different combinations of these three parameters produce identical solutions in the image plane, but geometrically scaled solutions
in the source plane, thus affecting the measurement of the source size and magnification. However, the same author also showed that the use of a randomly
initialized adaptive grid (the same adopted in this work), with a fixed number of degree-of-freedom, removes the biases associated with this degeneracy.
We will test our assumption of a SIE profile in a future paper using available HST and ALMA data, by comparing the lens modelling results obtained for $\alpha=2$
with those derived for a generic power-law model (Negrello et al. in prep.).\\
The SIE profile is described by 5 parameters: the displacement of the lens centroid, $\Delta x_{\rm L}$ and $\Delta y_{\rm L}$, with respect to the
centre of the image, the Einstein radius, $\theta_{\rm E}$, the minor-to-major axis ratio, $q_{\rm L}$, the orientation of the semi-major axis,
$\theta_{\rm L}$, measured counter-clockwise from West. For simplicity we do not include an external shear unless it is needed to improve the
modelling. In that case, its effect is described by two additional parameters: the shear strength, $\gamma$, and the shear angle, $\theta_{\gamma}$,
also measured counter-clockwise from West, thus raising the total number of free parameters from 5 to 7.\\

\subsection{Implementation}

The lens parameter space is explored using {\sc multinest} \citep{2008MNRAS.384..449F,2009MNRAS.398.1601F}, a Monte Carlo technique implementing the
nested sampling described in \cite{skilling2006}. Flat priors are adopted for the lens model, within the range: 
$0.1\,{\rm arcsec} \leq \theta_{\rm E} \leq 3.0\,{\rm arcsec}$; 
$0^{\circ} \leq \theta_{\rm L} < 180^{\circ}$; 
$0.2 \leq q_{\rm L} < 1.0$;
$-0.5\,{\rm arcsec}\leq \Delta x_{\rm L} \leq 0.5\,{\rm arcsec}$;
$-0.5\,{\rm arcsec}\leq \Delta y_{\rm L} \leq 0.5\,{\rm arcsec}$;
$0.0 \leq \gamma \leq 0.3$;
$0^{\circ} \leq \theta_{\gamma} < 180^{\circ}$.
In order to lighten the computational effort, a mask is applied to the IP pixels, keeping only those relevant, i.e. containing the lensed image, with
minimum background sky. These are then traced back to the SP where they define the area used for the source reconstruction.\\
As suggested in N15, a nuisance in lens modeling algorithms is the existence of unrealistic solutions, occupying significant regions of the
parameter space where the Monte Carlo method gets stuck. In general these local minima of the evidence correspond to a reconstructed SP that 
resembles a demagnified version of the observed IP. In order to avoid them, the first search of the parameter space is performed on a selected
grid of values of the free parameters, following the methods presented in N15. Then, the regions occupied by unrealistic solutions are excluded from
the subsequent search. Once the best lens model parameters are identified, a final {\sc multinest} run is employed to sample the posterior
distribution function (PDF), and to estimate the corresponding uncertainties, which are quoted as the 16th and 84th percentile of the PDF. \\
A fundamental quantity provided by the lens modelling is the magnification factor, $\mu$. This is defined as the ratio between the total flux
density of the source, as measured in the SP, and the total flux density of the corresponding images in the IP. In practice we estimate it as 
$\mu=F^{\rm IP}_{N\sigma}/F^{\rm SP}_{N\sigma}$ where $F^{\rm SP}_{N\sigma}$ is the flux density contributed by all the pixels in the SP with 
signal-to-noise ratio ${\rm SNR}\geq N$, while $F^{\rm IP}_{N\sigma}$ is the summed flux density of the all pixels within the corresponding region 
in the IP. We compute the value of $\mu$ for $N=3$ and $N=5$, taking the latter as our reference case. The uncertainty on the magnification factor 
is derived by calculating $\mu$ 1000 times, each time perturbing the lens model parameters around their best-fitting values; the final magnification 
factor is the median of the resulting distribution with errors given by the 16th and 84th percentile of the same distribution.

\section{Results and discussion}\label{sec:results}

The best-fitting values of the lens model parameters are reported in Table\,\ref{table:parameters}, while the results of the source reconstruction are 
shown in Fig.\,\ref{fig:lens_modelling_results}. The first panel on the left is the SMA dirty image, generated by adopting a natural weighting 
scheme. The second and the third panels from the left show the reconstructed IP and the residuals, respectively. The latter are derived by 
subtracting the model visibilities from the observed ones and then imaging the differences. The panel on the right shows the reconstructed source 
with contours at 3$\sigma$ (black curve) and 5$\sigma$ (white curves), while the second panel from the right shows the image obtained by assuming 
the best-fitting lens model and performing the gravitational lensing directly on the reconstructed source. The lensed image obtained in this way is 
unaffected by the sampling of the $uv$ plane and can thus help to recognize in the SMA dirty image those features that are really associated 
with the emission from the background galaxy. 

The estimated magnification factors, $\mu_{3\sigma}$ and $\mu_{5\sigma}$, are listed in Table\,\ref{tab:source_properties1} for the two adopted
values of the signal-to-noise ratios in the SP, i.e. SNR$\,\geq3$ and SNR$\,\geq5$, respectively. The area, $A_{\rm dust}$, of the regions in the 
SP used to compute the magnification factors is also listed in the same table together with the corresponding {\it effective radius}, $r_{\rm eff}$.
The latter is defined as the radius of a circle of area equal to $A_{\rm dust}$. We note that, despite the difference in the value of the area in 
the two cases, the derived magnification factors are consistent with each other. In fact, as the area decreases when increasing the SNR from 3
to 5, the centre of the selected region, in general, moves away from the caustic, where the magnification is higher. The two effects tend to 
compensate each other, thus reducing the change in the total magnification. Below we discuss our findings with respect to the results of
B13 and other results from the literature.

\begin{figure*}
\includegraphics[width=18.5cm]{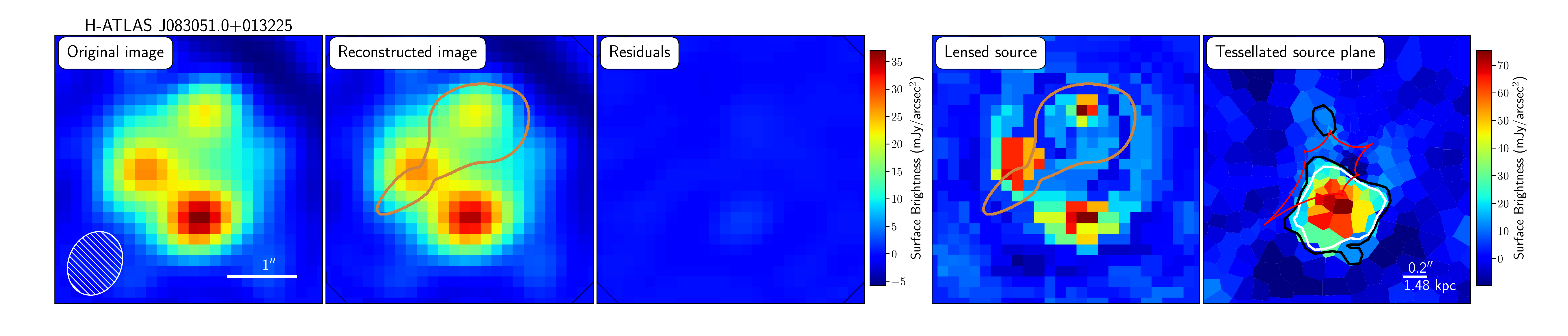} \\
\vspace{-0.5cm}
\includegraphics[width=18.5cm]{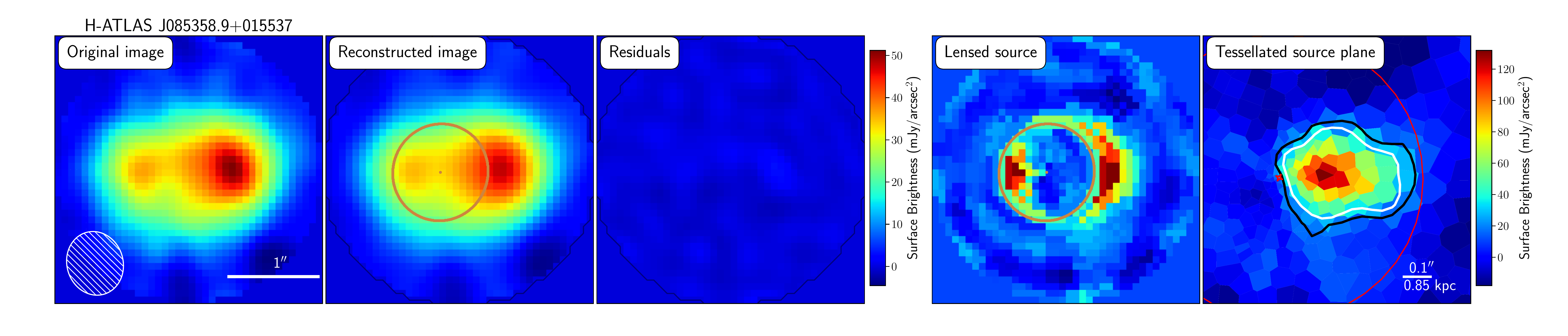}\\
\vspace{-0.5cm}
\includegraphics[width=18.5cm]{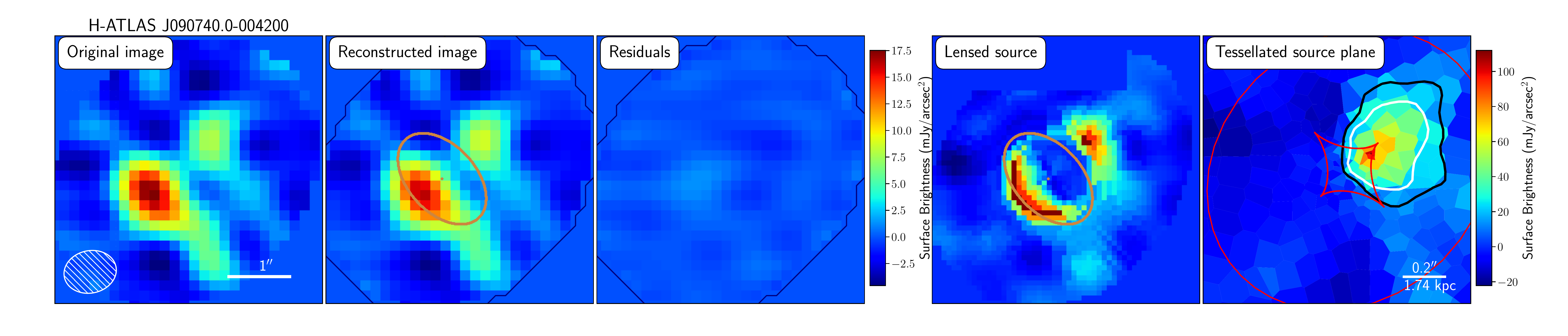} \\
\vspace{-0.5cm}
\includegraphics[width=18.5cm]{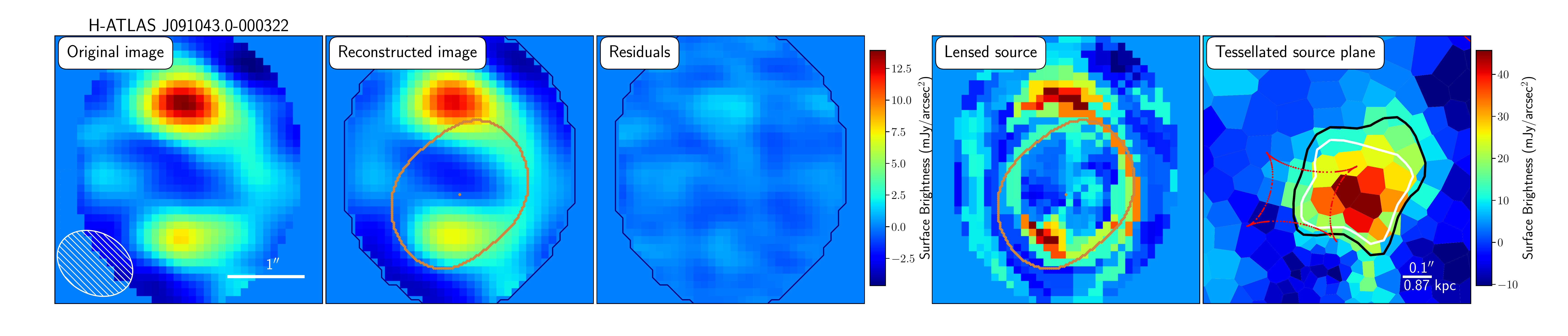} \\
\vspace{-0.5cm}
\includegraphics[width=18.5cm]{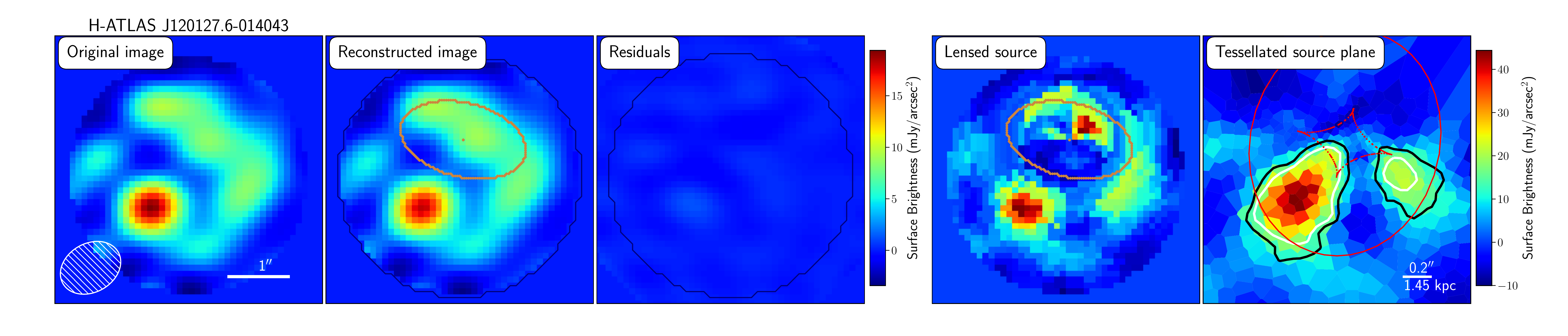} \\
\vspace{-0.5cm}
\includegraphics[width=18.5cm]{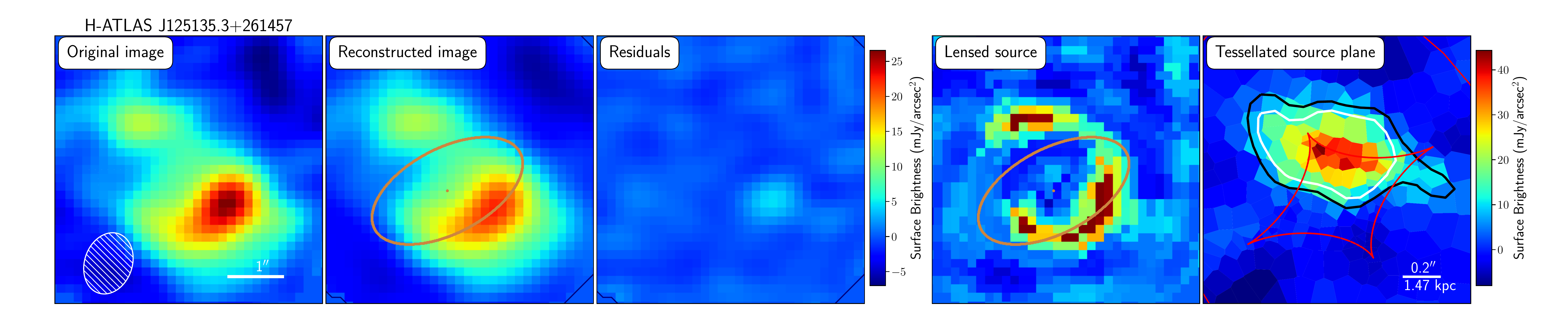} 
\vspace{-0.5cm}
\caption{
  Results of the lens modelling and source reconstruction. From left to right: input SMA image (created using a natural weighting scheme); 
  minimum $\chi^2$ image; residuals obtained by first subtracting the observed visibilities with the model ones and then transforming back to the 
  real space; image obtained by lensing the reconstructed source plane using the best-fitting lens model; the reconstructed background source with 
  contours at 3$\,\sigma$ (black curve) and 5$\,\sigma$ (white curve). The caustics and the critical lines are shown in brown (in the second and
  fourth panels from left) and in red (in the right panel), respectively. The white hatched ellipse in the bottom left corner of the leftmost 
  panels represents the SMA synthesized beam. }
  \label{fig:lens_modelling_results}
     \addtocounter{figure}{-1}
\end{figure*}

\begin{figure*}
\includegraphics[width=18.5cm]{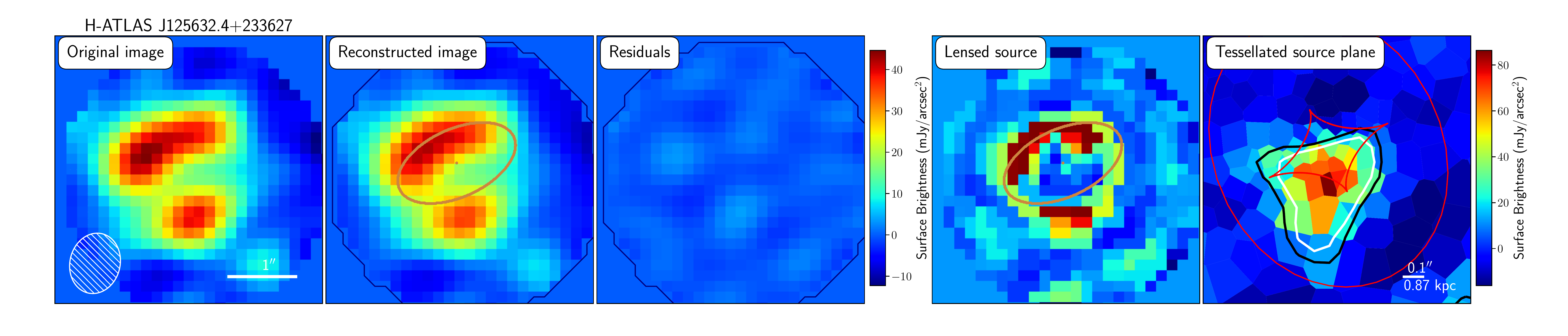} \\
\vspace{-0.5cm}
\includegraphics[width=18.5cm]{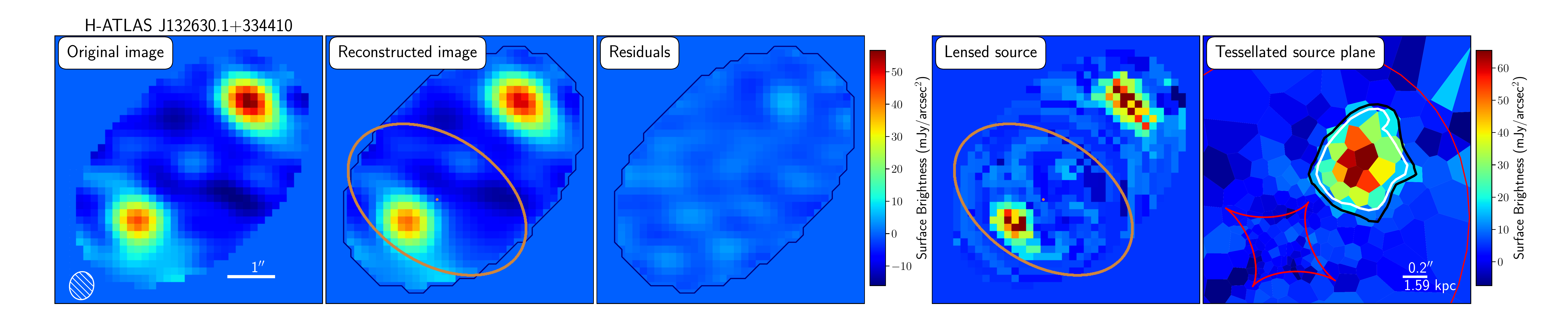} \\
\vspace{-0.5cm}
\includegraphics[width=18.5cm]{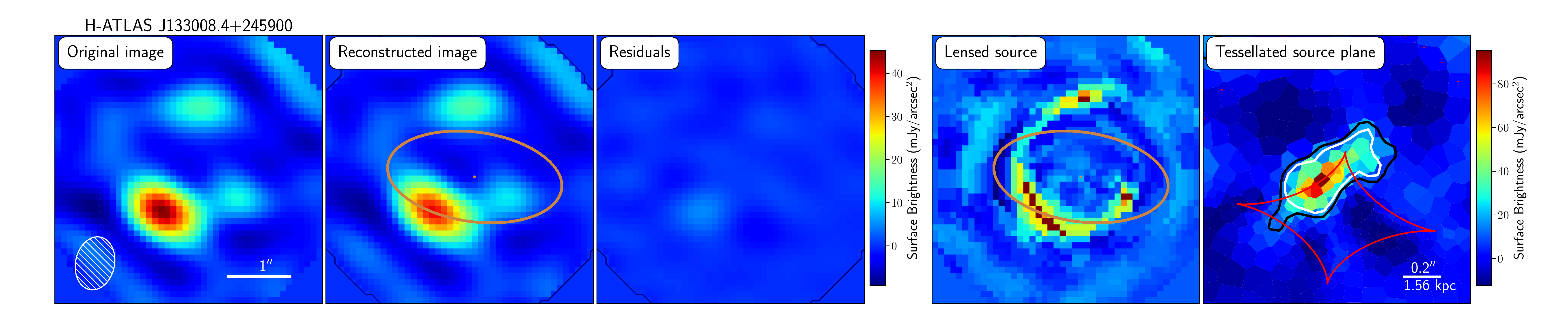} \\
\vspace{-0.5cm}
\includegraphics[width=18.5cm]{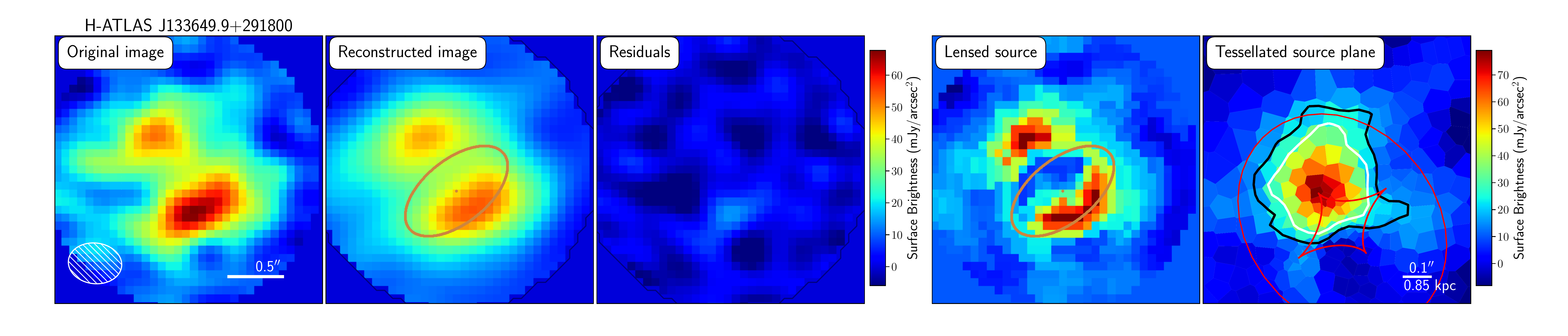} \\
\vspace{-0.5cm}
\includegraphics[width=18.5cm]{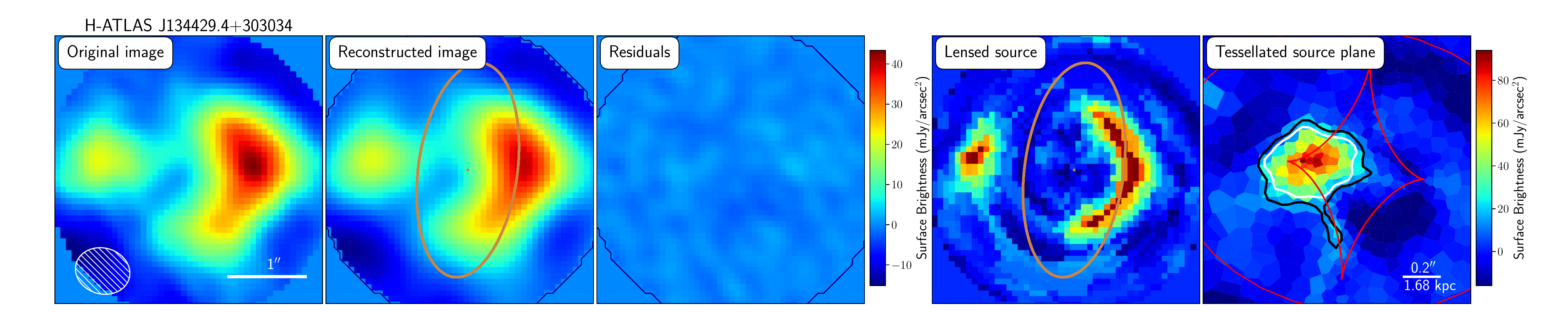} \\
\vspace{-0.5cm}
\includegraphics[width=18.5cm]{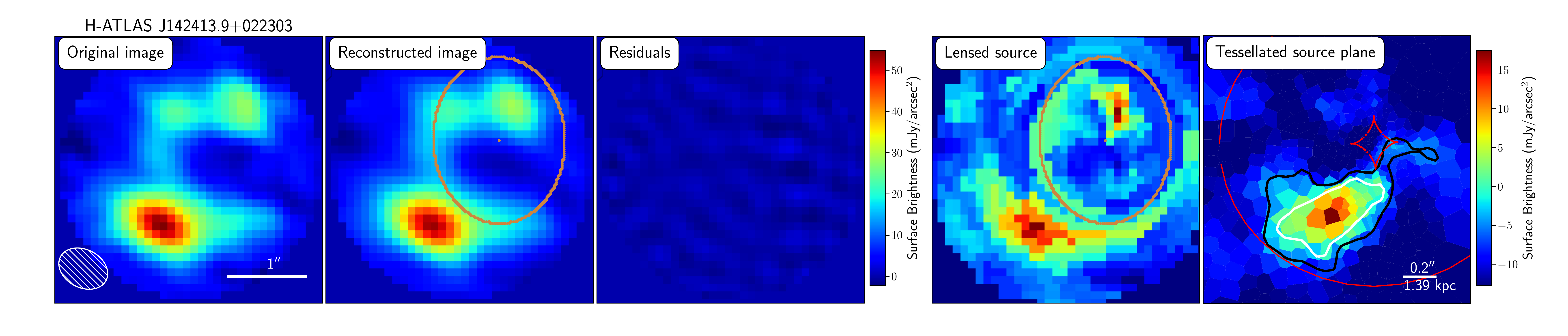}
\vspace{-0.5cm}
\caption{\it $-$ continued}
\end{figure*}

\begin{figure*}
\begin{centering}
\includegraphics[width=\textwidth]{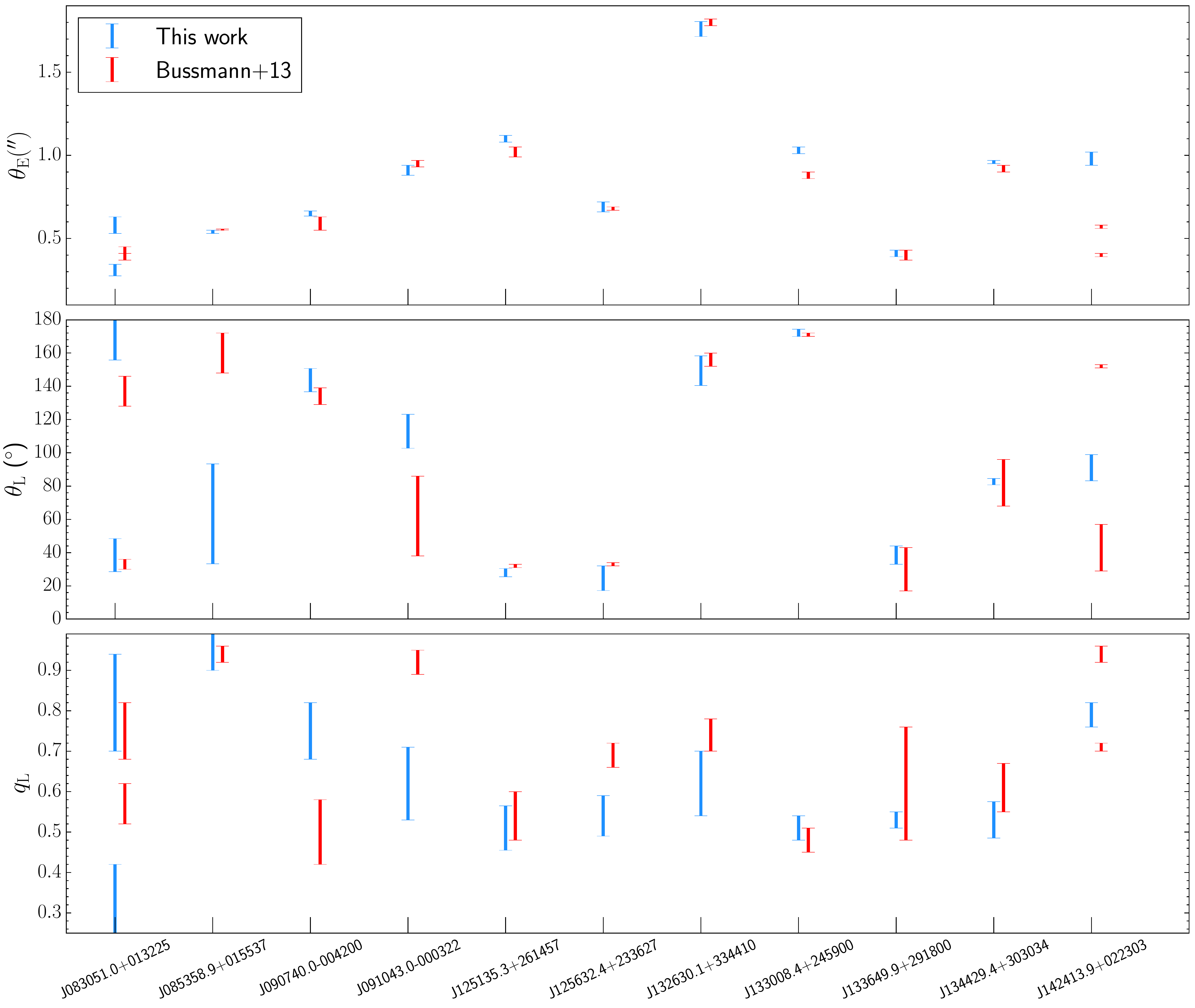}
\end{centering}
\caption{
  Comparison between our results (blue error bars) and those of B13 (red error bars) for the parameters of the SIE lens mass model: Einstein
  radius, $\theta_{\rm E}$, rotation angle, $\theta_{\rm L}$, and minor-to-major axis ratio, $q_{\rm L}$. Two datapoints are plotted whenever
  two lenses are employed in lens modeling (HATLASJ083051.0+013225, HATLASJ142413.9+022303).
  }
\label{fig:Busscomp}
\end{figure*}

\begin{table*}
\centering
\caption{
  Lens modeling results: source properties. Magnifications, $\mu_{3\sigma}$ and $\mu_{5\sigma}$, are evaluated as the ratio between the total
  flux density of the region in the SP with SNR$\ge3$ and SNR$\ge5$, respectively, and the total flux density of the corresponding region in the
  IP. $A_{\rm dust, 3\sigma}$ and $A_{\rm dust, 5\sigma}$ are the areas of the regions with SNR$\ge3$ and SNR$\ge5$ in the source plane, while
  $r_{\rm eff, 3\sigma}$ and $r_{\rm eff, 5\sigma}$, are the radius of a circle with area equal to $A_{\rm dust, 3\sigma}$ and $A_{\rm dust, 5\sigma}$,
  respectively. FHWMs are the values of the FWHM of the major and minor axis length obtained from the Gaussian fit to the reconstructed source
  surface brightness, while FWHM$_{\rm m}=\sqrt{{\rm FWHM_{maj}\times FWHM_{min}}}$.
  }
\label{tab:source_properties1}
\begin{tabular}{lccccccccc}
\hline
{\it H}-ATLAS IAU name  &$\mu_{3\sigma}$ &$\mu_{5\sigma}$ & $A_{\rm dust, 3\sigma}$ & $A_{\rm dust, 5\sigma}$ & $r_{\rm eff, 3\sigma}$ & $r_{\rm eff, 5\sigma}$ & FWHMs & FWHMm \\
    &   &   & (kpc$^2$)                & (kpc$^2$)          & (kpc)             & (kpc)                & (kpc) & (kpc) \\
\hline
HATLASJ083051.0+013225  & 4.25$\pm$0.68 & 4.04$\pm$0.70 & 33.5$\pm$5.6  & 22.3$\pm$2.8  & 3.27$\pm$0.27 & 2.67$\pm$0.17 & 1.64/1.46 & 1.54$\pm$0.10 \\
HATLASJ085358.9+015537  & 5.40$\pm$1.76 & 5.26$\pm$1.82 & 16.1$\pm$6.0  &  9.5$\pm$3.7  & 2.26$\pm$0.41 & 1.74$\pm$0.34 & 1.72/1.14 & 1.37$\pm$0.33 \\
HATLASJ090740.0$-$004200& 6.73$\pm$0.93 & 7.51$\pm$1.31 & 18.4$\pm$4.1  & 10.2$\pm$2.0  & 2.42$\pm$0.27 & 1.80$\pm$0.19 & 1.83/1.24 & 1.46$\pm$0.18 \\
HATLASJ091043.1$-$000321& 6.63$\pm$0.68 & 6.89$\pm$0.79 & 10.6$\pm$2.7  &  5.5$\pm$1.6  & 1.84$\pm$0.23 & 1.33$\pm$0.20 & 1.32/1.17 & 1.24$\pm$0.19 \\
HATLASJ120127.6$-$014043& 3.30$\pm$0.55 & 3.03$\pm$0.59 & 33.3$\pm$4.8  & 17.4$\pm$2.4  & 3.25$\pm$0.23 & 2.36$\pm$0.16 & 1.87/1.33 & 1.57$\pm$0.07 \\
HATLASJ125135.4+261457  & 8.38$\pm$0.54 & 9.16$\pm$0.78 & 23.3$\pm$2.9  & 15.1$\pm$2.1  & 2.72$\pm$0.17 & 2.19$\pm$0.15 & 2.19/1.12 & 1.56$\pm$0.08 \\
HATLASJ125632.7+233625  & 5.90$\pm$1.29 & 6.85$\pm$1.67 & 22.3$\pm$5.6  & 11.1$\pm$3.4  & 2.66$\pm$0.34 & 1.88$\pm$0.29 & 1.36/1.32 & 1.34$\pm$0.23 \\
HATLASJ132630.1+334410  & 3.20$\pm$0.57 & 3.24$\pm$0.54 & 41.6$\pm$8.4  & 29.9$\pm$5.9  & 3.64$\pm$0.36 & 3.09$\pm$0.34 & 2.06/1.67 & 1.86$\pm$0.17 \\
HATLASJ133008.4+245900  & 9.62$\pm$0.98 & 9.89$\pm$1.01 & 14.0$\pm$2.4  &  8.5$\pm$2.1  & 2.11$\pm$0.18 & 1.65$\pm$0.20 & 1.64/0.70 & 1.07$\pm$0.10 \\
HATLASJ133649.9+291801  & 4.79$\pm$0.37 & 5.34$\pm$0.56 & 14.2$\pm$1.6  &  7.3$\pm$1.0  & 2.13$\pm$0.12 & 1.52$\pm$0.10 & 1.57/1.40 & 1.48$\pm$0.09 \\
HATLASJ134429.4+303036  & 8.35$\pm$0.95 & 8.97$\pm$1.17 & 14.4$\pm$2.4  &  9.2$\pm$1.8  & 2.14$\pm$0.18 & 1.71$\pm$0.19 & 1.52/1.00 & 1.24$\pm$0.12 \\
HATLASJ142413.9+022303  & 4.21$\pm$0.69 & 3.69$\pm$0.47 & 19.7$\pm$2.4  &  9.1$\pm$1.5  & 2.51$\pm$0.14 & 1.71$\pm$0.14 & 2.04/1.11 & 1.50$\pm$0.12 \\
\hline
\end{tabular}
\end{table*}
\begin{figure*}
\begin{centering}
\includegraphics[width=\textwidth]{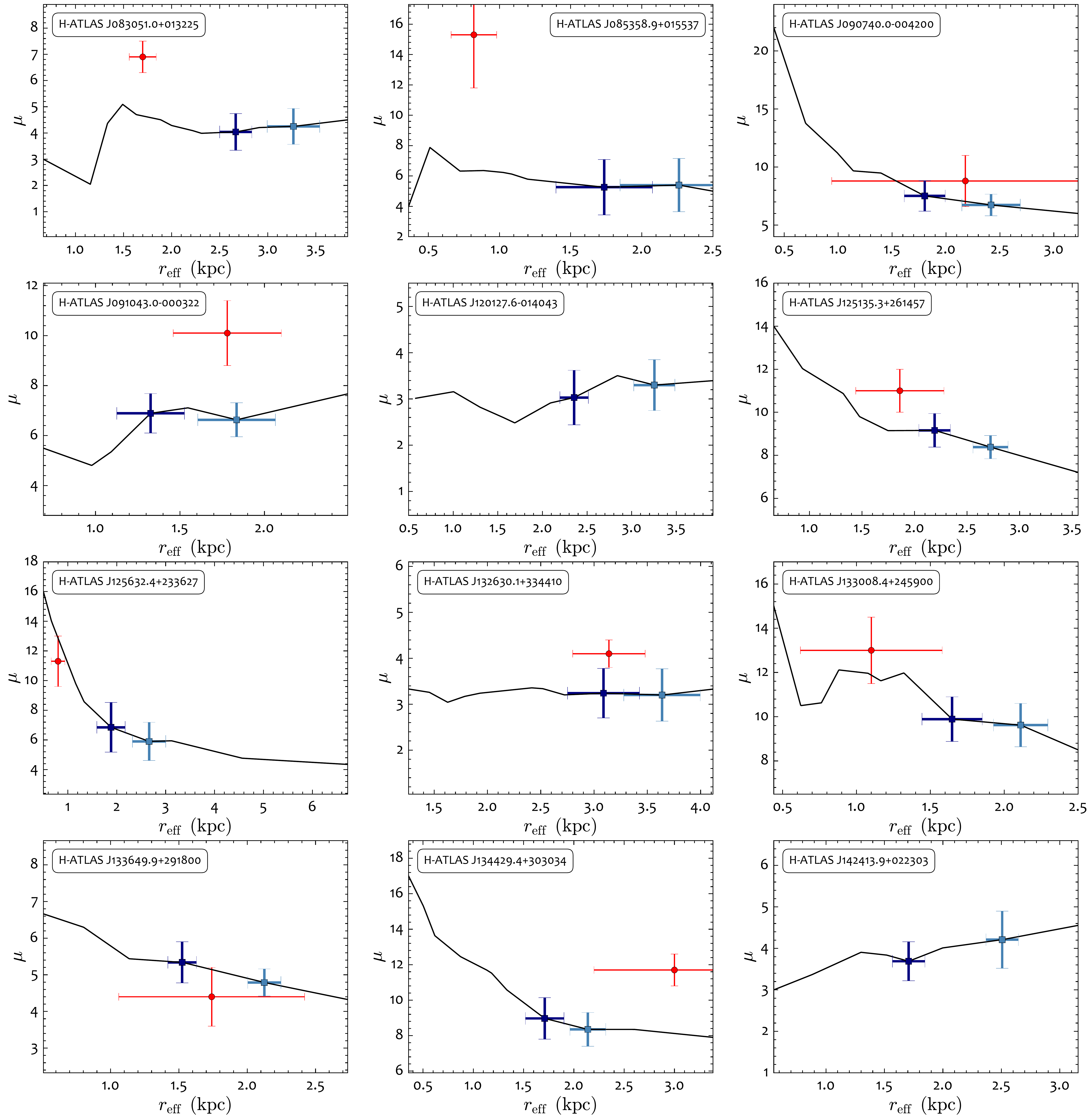}
\end{centering}
\vspace{-0.5cm}
\caption{
  Magnification profiles of the reconstructed sources. The magnification factor, $\mu$, is evaluated in steps of SNR in the SP, from two up to the
  maximum and shown as a function of the effective radius of the area defined by the SP pixels with SNR above the adopted steps. The squares mark
  the values of the magnification calculated for SNR$\,=3$ (outermost; light blue square) and SNR$\,=5$ (innermost; dark blue square). The red 
  point is the magnification factor estimated by B13. We have placed it at a radius corresponding to $2\times r_{\rm half}$, as this is the 
  radius of the region in the SP used by B13 to compute the magnification. For HATLASJ142313.9+022303, the point of B13 is located outside the 
  plotted region, at $r_{\rm eff}\sim7\,$kpc.}
  \label{fig:magn_profiles}
\end{figure*}

\subsection{Lens parameters}\label{subsec:lens_param}

Fig.\,\ref{fig:Busscomp} compares our estimates of the lens mass model parameters with those of B13.
In general we find good agreement, although there are some exceptions (e.g. HATLASJ133008.4+245900), particularly when multiple lenses are 
involved in the modelling, i.e. for HATLASJ083051.0+013225 and HATLASJ142413.9+022303. We briefly discuss each case individually.
\\

{\it HATLASJ083051.0+013225}: This is a relatively complex system (see Fig.\,1 of N17), with two foreground objects at different redshifts (B17) 
revealed at 1.1\,$\micron$ and 2.2\,$\micron$ by observations with {\it HST}/WFC3 (N17; Negrello et al. in prep.) and {\it Keck}/AO
\citep{2014ApJ...797..138C}, respectively. However the same data show some elongated structure north of the two lenses, which may be associated 
with the background galaxy, although this is still unclear due to the apparent lack of counter-images (a detailed lens modelling of this system 
performed on ALMA+{\it HST}+{\it Keck} data is currently ongoing; Negrello et al. in prep.). In our modelling we have assumed that the two lenses 
are at the same redshift, consistently with the treatment by B13. However, compared to B13, we derive an Einstein radius that is higher for one lens
($0.57^{\prime\prime}$ versus $0.43^{\prime\prime}$) and lower for the other one ($0.31^{\prime\prime}$ versus $0.39^{\prime\prime}$). The
discrepancy is likely due to the complexity of the system, which may induce degeneracies among the model parameters; however it is worth mentioning
that while we keep the position of both lenses as free parameters, B13 fixed the position of the second lens with respect to the first one, by
setting the separation between the two foreground objects equal to that measured in the near-IR image. 

{\it HATLASJ085358.9+015537}: This system was observed with {\it Keck}/NIRC2 in the $K_{s}$-band \citep{2014ApJ...797..138C}. 
The background galaxy is detected in the near-IR in the form of a ring-like structure that was modelled by Calanog et al. assuming a SIE model for 
the lens and a S{\'e}rsic profile for the background source surface brightness. Our modelling of the SMA data gives results for the lens mass model 
consistent with those of Calanog et al., both indicating an almost spherical lens. B13 also find a nearly spherical lens ($q_{\rm L}\sim0.94$) but 
with a different rotation angle ($\theta_{\rm L}\sim160^{\circ}$ versus $\theta_{\rm L}\sim62^{\circ}$), even though the discrepance is less than
3$\sigma$ once considered the higher confidence interval consequence of a spherical lens.

{\it HATLASJ090740.0$-$004200}: This is one of the first five lensed galaxies discovered in {\it H}-ATLAS \citep{2010Sci...330..800N}, and is also 
known as SDP.9. High-resolution observations at different wavelengths are available for this system, from the near-IR with {\it HST}/WFC3
\citep{2014MNRAS.440.1999N}, to sub-mm with NOEMA \citep{2017arXiv170105901O} or 1.1mm with ALMA \citep{2017ApJ...843L..35W}, to the X-ray band with
Chandra \citep{2017arXiv170910427M}. The results of our lens modelling of the SMA data are consistent with those obtained by other groups
at different wavelengths \citep[e.g.]{2014MNRAS.440.2013D, 2017arXiv170910427M}. However, B13 found a significantly lower lens axis-ratio
compared to our estimate ($q_{\rm L}=0.50$ versus $q_{\rm L}=0.75$).

{\it HATLASJ091043.1$-$000321}: This is SDP.11, another of the first lensed galaxies discovered in {\it H}-ATLAS \citep{2010Sci...330..800N}.
{\it HST}/WFC3 imaging data at 1.1$\micron$ reveals an elongated Einstein ring \citep{2014MNRAS.440.1999N}, hinting to the effect of an external shear 
possibly associated with a nearby edge-on galaxy. In fact, \cite{2014MNRAS.440.2013D} introduced an external shear in their lens modelling of this 
system, which they constrained to have strength $\gamma\sim0.23$. We also account for an external shear in our analysis. Our results are consistent
with those of Dye et al. They also agree with the Einstein radius estimated by B13, although our lens is significantly more elongated and has a 
higher rotation angle. It is worth noticing, though, that B13 does not introduce an external shear in their analysis, which may explain the 
difference in the derived lens axial ratio.

{\it HATLASJ120127.5$-$014043}: This is the {\it H}-ATLAS source that we have confirmed to be lensed with the new SMA data. It is the only object 
in our sample for which we still lack a spectroscopic measure of the redshift of the background galaxy. The redshift estimated from the
{\it Herschel}/SPIRE photometry is $z_{\rm sub-mm}=3.80\pm0.58$. The reconstructed source is resolved into two knots of emission, separated by
$\sim3.5\,$kpc. 

{\it HATLASJ125135.4+261457}: The estimated Einstein radius is slightly higher than reported by B13
($\theta_{\rm E}=1.10\pm0.02^{\prime\prime}$ versus $\theta_{\rm E}=1.02\pm0.03^{\prime\prime}$) while the rotation angle of the lens is smaller
($\theta_{\rm L}=28\pm2.5^{\circ}$ versus $\theta_{\rm L}=38\pm1^{\circ}$). The reconstructed source is quite elongated, extending in the SW to NE
direction, with a shape that deviates from a perfect ellipse. This might suggest that, at the scale probed by the SMA observations, the source 
comprises two partially blended components. This morphology is not accounted for by a single elliptical S{\'e}rsic profile, which may explain the
observed discrepancies with the results of B13.

{\it HATLASJ125632.7+233625}: For this system we find a lens that is more elongated compared to the value derived by B13 ($q_{\rm L}=0.54\pm0.09$ versus
$q_{\rm L}=0.69\pm0.03$). The reconstructed source morphology has a triangular shape which may bias the results on the lens parameters when the 
modelling is performed under the assumption of a single elliptical S{\'e}rsic profile, as in B13. 

{\it HATLASJ132630.1+334410}: The background galaxy is lensed into two images, separated by $\sim3.5^{\prime\prime}$, none of them resembling an 
arc. This suggests that the source is not lying on top of the tangential caustic, but away from it, although still inside the radial caustic to
account for the presence of two images. As revealed by {\it HST}/WFC3 observations (see N17, their Fig.\,3), the lens is located close to the 
southernmost lensed image. The lack of extended structures, like arcs or rings, makes the lens modelling more prone to degeneracies. Despite that, 
we find a good agreement with the results of B13.

{\it HATLASJ133008.4+245900}: Besides the lens modelling performed by B13 on SMA data, this system was also analysed by \cite{2014ApJ...797..138C}
using {\it Keck}/AO $K_{s}$-band observations, where the background galaxy is detected. The configuration of the multiple images is similar in the
near-IR and in the sub-mm suggesting that the stellar and dust emission are co-spatial. We derive an Einstein radius
$\theta_{\rm E}=1.03^{\prime\prime}$, higher than B13's result ($\theta_{\rm E}=0.88^{\prime\prime}$). Our estimate is instead in agreement
with the finding of \cite{2014ApJ...797..138C} and Negrello et al. (in prep.; based on  {\it HST}/WFC3 imaging data). Interestingly, the 
reconstructed background source is very elongated. This is due to the presence of two partially blended knots of emission, a main one extending 
across the tangential caustic and a second, fainter one located just off the fold of the caustic. This is another example where the assumption 
that the source is represented by a single S{\'e}rsic profile, made by B13, is probably affecting the estimated lens model parameters.

{\it HATLASJ133649.9+291801}: This is the single-lens system in our sample with the smallest Einstein radius, $\theta_{\rm E}=0.4^{\prime\prime}$.
Our lens modelling gives results consistent with those of B13.

{\it HATLASJ134429.4+303036}: This is the 500$\,\mu$m brightest lensed galaxy in the entire N17 sample. The observed lensed images indicate a typical
cusp configuration, similar to what was observed in the well studied lensed galaxy  SDP.81 \citep[e.g.][]{2015MNRAS.452.2258D}, where the background
galaxy lies on the fold of the tangential caustic. According to our modelling the lens is significantly elongated ($q_{\rm L}=0.53$) in the North-South
direction, consistent with what is indicated by available {\it HST}/WFC3 imaging data for the light distribution of the foreground galaxy
(see Fig.\,3 of N17). We estimate a higher Einstein radius than the one reported by B13, although the two results are still consistent within 2$\sigma$.

{\it HATLASJ142413.9+022303}: This source $-$ a 500$\,\mu$m ``riser'' $-$ was first presented in \cite{2011ApJ...740...63C} while the lens 
modelling, based on SMA data, was performed in \cite{2012ApJ...756..134B}. Observations carried out with {\it HST}/WFC3 and {\it Keck}/AO
\citep{2014ApJ...797..138C} revealed two foreground galaxies, separated by $\sim0.3^{\prime\prime}$, although only one currently has a 
spectroscopic redshift, $z=0.595$ (B13). No emission from the background galaxy is detected in the near-IR. B13 modelled the system using two SIE
profiles. We attempted the same but found no significant improvement in the results compared to the case of a single SIE mass distribution, 
which we have adopted here. We find $\theta_{\rm E}=0.97^{\prime\prime}$, consistent with the value derived from the lens modelling of ALMA
data performed by \cite{2017arXiv170505413D}, which also assumed a single SIE profile. On the other hand B13 obtained
$\theta_{\rm E,1}=0.57^{\prime\prime}$ and $\theta_{\rm E,2}=0.40^{\prime\prime}$ for the two lenses. In this case the comparison with the B13 
results is not straightforward. It is also important to note, as shown in Fig.\,\ref{fig:lens_modelling_results}
[see also \cite{2017arXiv170505413D}], that the background source has a complex extended morphology, which cannot be recovered by a single S{\'e}rsic
profile. \cite{2012ApJ...756..134B} modelled this system assuming two S{\'e}rsic profiles for the background galaxy but their results, particularly for 
the position of the second knot of emission in the SP, disagree with ours and with the findings of \cite{2017arXiv170505413D}.\\

According to our findings, in single-lens systems the use of an analytic model for the source surface brightness does not bias the results on the SIE
lens parameters as long as the background galaxy is not partially resolved into multiple knots of emission. A way to overcome this problem would be
to test the robustness of the results by adding a second source during the fitting procedure. However, the drawback of this approach is the 
increase in the number of free parameters, and, therefore, the increased risk of degeneracies in the final solution. We conceived our SLI method
to overcome this problem and we recommend it in the modelling of lensed galaxies\footnote{The codes used here are available upon request
but will be made soon available via GitHub and at the webpage http://www.mattianegrello.com}.

\begin{figure}
\begin{centering}
\includegraphics[width=\columnwidth]{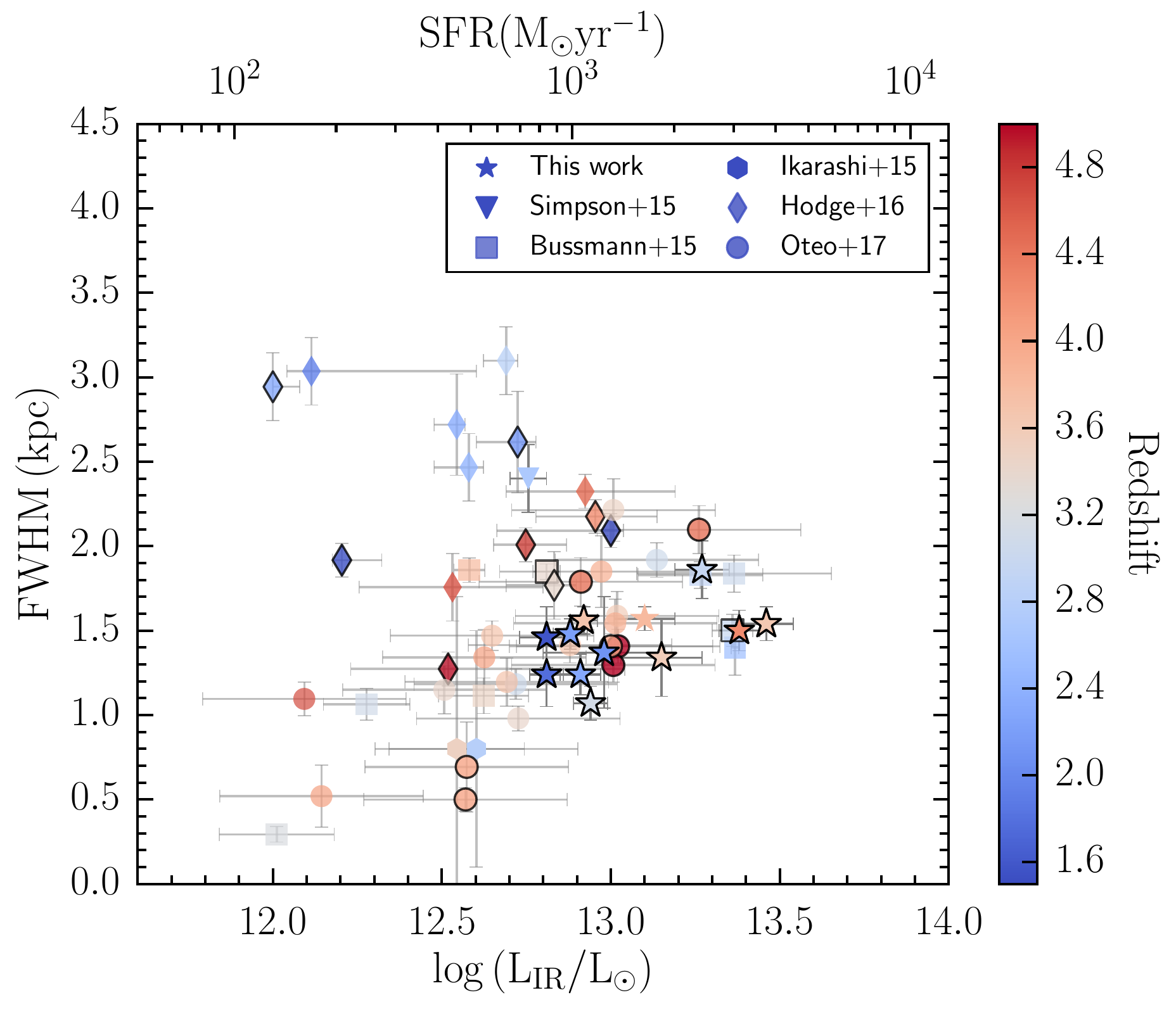}
\end{centering}
\vspace{-0.5cm}
\caption{
    FWHMs of the sources in our sample (stars) as a function of their infrared luminosity, compared to results from literature:
  \citet[][triangle]{2015ApJ...807..128S}, \citet[][squares]{2015ApJ...812...43B}, \citet[][hexagons]{2015ApJ...810..133I},
  \citet[][diamonds]{2016ApJ...833..103H}, \citet[][circles]{2017arXiv170904191O}.
  All the sources taken from literature were fitted or (for the lensed ones) modelled using Gaussians and here we report,
  as an effective radius, the geometric mean of the values of the FWHM along the minor and the major axis. The only exception
  is the point of Simpson et al. which represents the median of the FWHM$_{\rm major}$ values for their sample (see
  Section\,\ref{subsec:magn_sizes}). The data points are colored according to their redshift. Most objects have a photometric
  redshift estimate; those with a spectroscopic redshift measurement are highlighted by a dense black outline.}\label{fig:reff_vs_Lir}
\end{figure}
\begin{figure*}
\begin{centering}
\includegraphics[width=\textwidth]{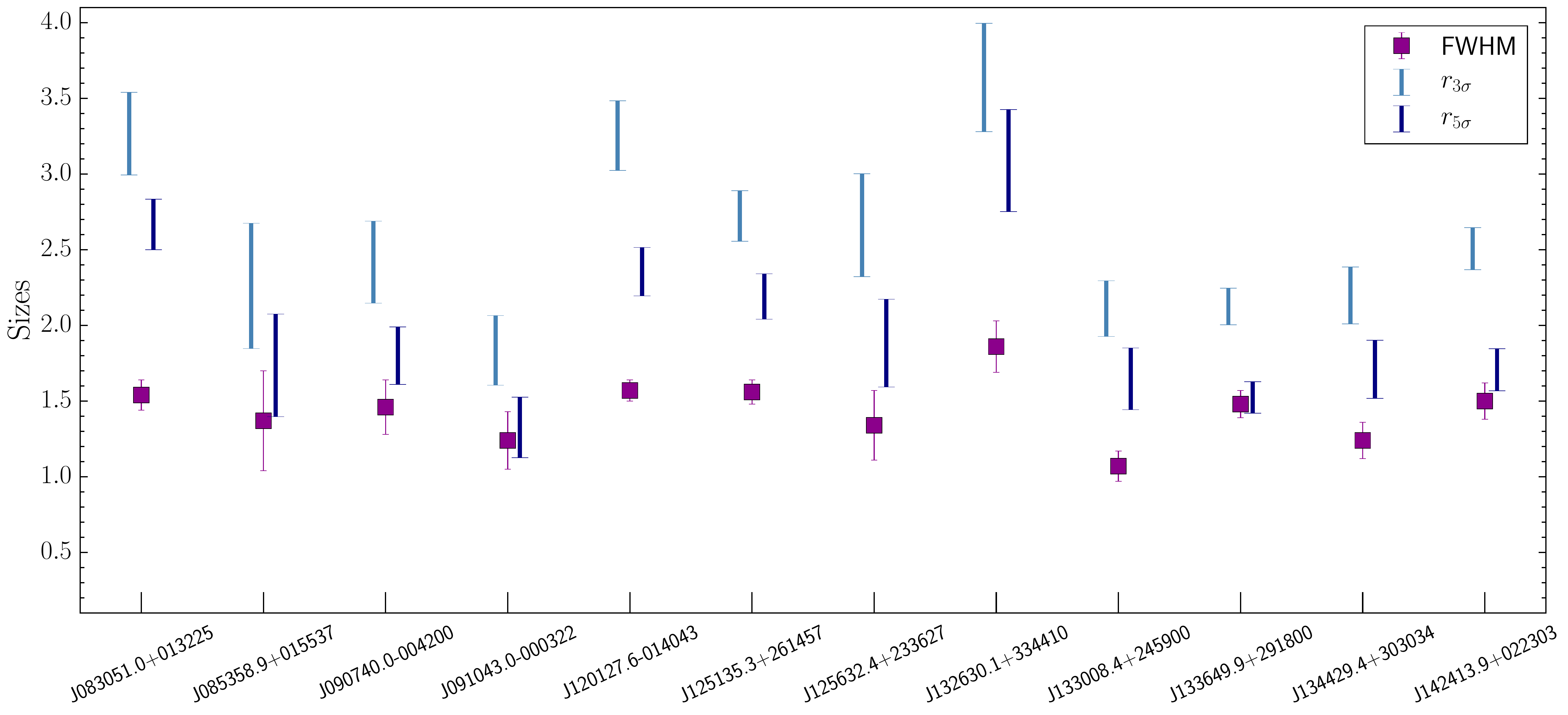}
\end{centering}
\vspace{-0.5cm}
\caption{
  Comparison between the measured $5\sigma$ and $3\sigma$ effective radius and the geometric mean of the values of the FWHM along the minor and
  the major axis obtained from a Gaussian fit to the reconstructed source plane. The FWHMs are systematically lower then the reported
  effective radii obtained over a certain value of signal-to-noise value, due to lost features not retrieved by the Gaussian fit.}\label{fig:reff_vs_FWHM}
\end{figure*}
\begin{figure*}
\begin{centering}
\includegraphics[width=\textwidth]{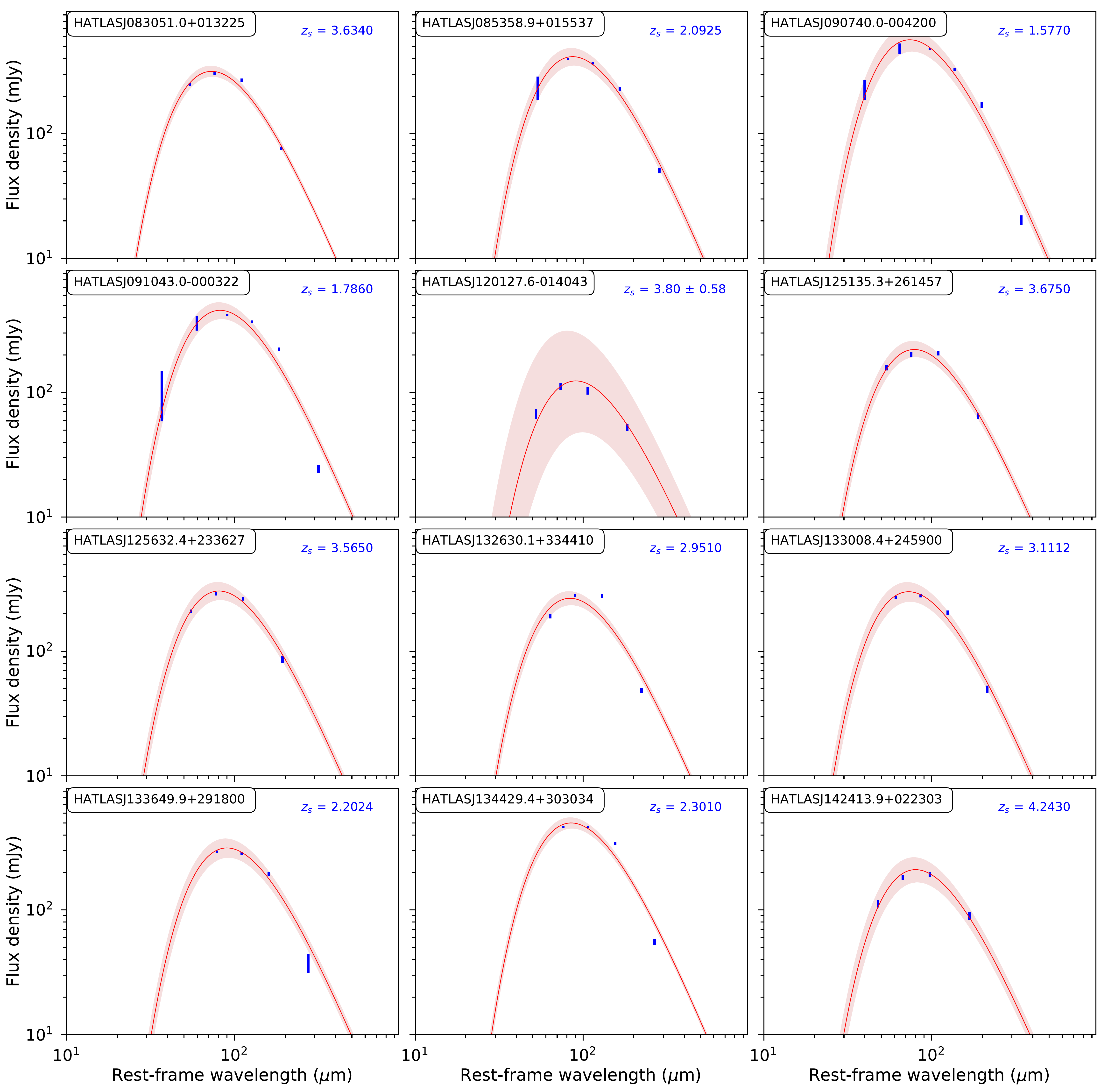}
\end{centering}
\vspace{-0.5cm}
\caption{
  Observed far-IR to sub-mm SEDs of the 12 sources (blue error bars; from {\it Herschel} and SMA) together with the best-fitting modified black-body 
  spectrum (red curve; assuming dust emissivity index $\beta = 1.5$). The shaded red area shows the 68\% confidence region associated with 
  the best-fitting model. The redshift of the source is reported on the top-right corner of each panel. The redshift is spectroscopic for all the 
  sources but one, i.e. HATLAS J120127.6-014043. This accounts for the significantly larger uncertainity in the fit to the SED of HATLAS 
  J120127.6-014043 compared to the other sources.}\label{fig:SEDs}
\end{figure*}
\begin{figure}
\begin{centering}
\includegraphics[width=8.3cm]{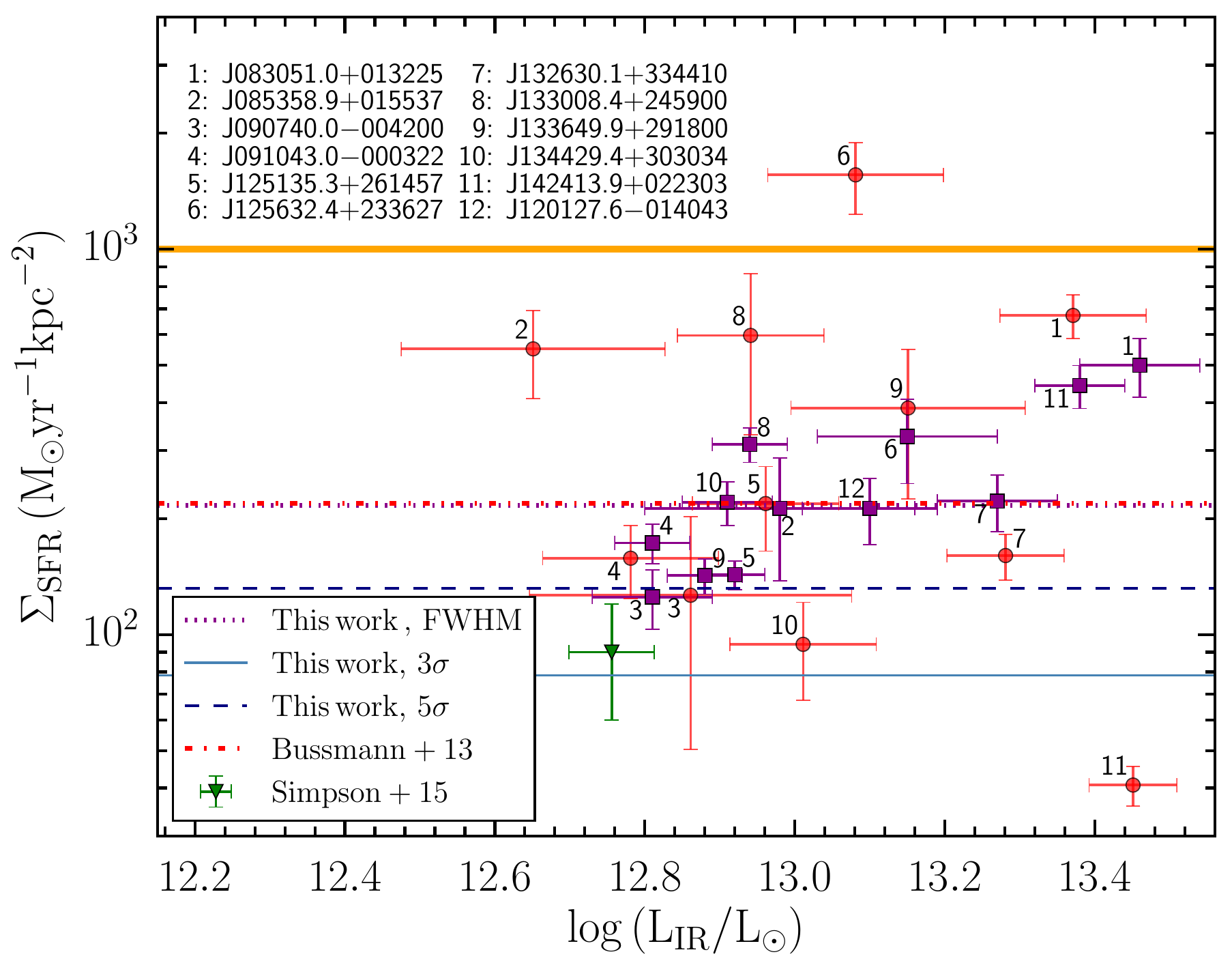}
\end{centering}
\vspace{-0.3cm}
\caption{
  Star formation rate surface density as a function of infrared luminosity for the sources in our sample. The dark magenta squares correspond to the
  case $\protect\mu=\protect\mu_{5\protect\sigma}$ and $A_{\rm dust}=A_{\rm dust,\,FWHM}$. The dotted dark magenta line marks their median value. The median value of
  $\protect\Sigma_{\rm SFR}$ for the case $\protect\mu=\protect\mu_{5\protect\sigma}$ and $A_{\rm dust}=A_{\rm dust,5\protect\sigma}$ is shown by the dotted blue line, while the case of
  $\protect\mu=\protect\mu_{3\protect\sigma}$ and $A_{\rm dust}=A_{\rm dust,3\protect\sigma}$ is the continuous cyan line. For comparison we show the data points from B13 (red dots;
  calculated as explained in Section\,\protect\ref{subsec:Lir}) and the median value of the SFR surface density of DSFGs from the \protect\cite{2015ApJ...807..128S}
  sample (green triangle). The yellow line marks the Eddington limit for a radiation pressure supported starburst galaxy \protect\citep{2005ApJ...630..167T, 2011EAS....52..275A}.}\label{fig:SigmaSFR_vs_Lir}
\end{figure}
\begin{table*}
\centering
\caption{Intrinsic properties of the 12 sources in our sample. The correction for the effect of lensing has been implemented by using the value
  $\protect\mu=\protect\mu_{3\protect\sigma}$ for the magnification as reported in Table\,\protect\ref{tab:source_properties1}.\\
  The dust temperature, ${\rm T}_{\rm dust}$, and the dust luminosities, ${\rm L}_{\rm FIR}$ (integrated in the rest-frame wavelength range $40-120\,\protect\mu$m) 
  and ${\rm L}_{\rm IR}$ (integrated over $8-1000\,\protect\mu$m in the rest-frame), are derived by fitting the {\it Herschel} and SMA photometry with a 
  modified blackbody spectrum with dust emissivity index $\protect\beta =1.5$, as in Bussmann et al. (2013). Star formation rates, SFR, are estimated 
  from ${\rm L}_{\rm IR}$ following \protect\cite{2012ARAA..50..531K}. The dust luminosity and SFR densities are computed as
  $\protect\Sigma_{\rm FIR}={\rm L}_{\rm FIR}/A_{\rm dust, 3\protect\sigma}$, $\protect\Sigma_{\rm IR}={\rm L}_{\rm IR}/A_{\rm dust, 3\protect\sigma}$ and
  $\protect\Sigma_{\rm SFR}={\rm SFR}/A_{\rm dust, 3\protect\sigma}$, with the values of $A_{\rm dust, 3\protect\sigma}$ taken from Table\,\protect\ref{tab:source_properties1}.}\label{tab:source_properties_3sigma}
\small
\begin{tabular}{lcccccccc}
\hline
IAUname         & ${\rm T}_{\rm dust}$    & $\log{{\rm L}_{\rm FIR}}$     & $\log{{\rm L}_{\rm IR}}$  & SFR     & $\log{\Sigma_{\rm FIR}}$   & $\log{\Sigma_{\rm IR}}$  & $\Sigma_{\rm SFR}$  \\
     & K      & (${\rm L}_{\sun}$)       & (${\rm L}_{\sun}$)     & (${\rm M}_{\sun}{\rm yr}^{-1}$)   & (${\rm L}_{\sun}$\,kpc$^{-2}$) & (${\rm L}_{\sun}$\,kpc$^{-2}$) & (${\rm M}_{\sun}{\rm yr}^{-1}$\,kpc$^{-2}$) \\
\hline          
HATLASJ083051.0+013225  & 44.4$\pm$0.6  & 13.15$\pm$0.08  & 13.44$\pm$0.08 & 3540$\pm$560 & 11.63$\pm$0.08 & 11.91$\pm$0.08 & 105$\pm$25 \\
HATLASJ085358.9+015537  & 37.4$\pm$0.7  & 12.94$\pm$0.17  & 12.97$\pm$0.17 & 1220$\pm$400 & 11.73$\pm$0.16 & 11.77$\pm$0.16 &  75$\pm$34 \\
HATLASJ090740.0$-$004200 & 43.9$\pm$1.2 & 12.60$\pm$0.06  & 12.86$\pm$0.06 &  940$\pm$130 & 11.34$\pm$0.08 & 11.59$\pm$0.08 &  51$\pm$11 \\
HATLASJ091043.1$-$000321 & 39.4$\pm$0.9 & 12.73$\pm$0.05  & 12.82$\pm$0.05 &  860$\pm$90  & 11.71$\pm$0.13 & 11.80$\pm$0.13 &  81$\pm$22 \\
HATLASJ120127.6$-$014043 & 35.9$\pm$3.9 & 13.07$\pm$0.08  & 13.07$\pm$0.08 & 1530$\pm$260 & 11.55$\pm$0.07 & 11.55$\pm$0.07 &  46$\pm$10 \\
HATLASJ125135.4+261457  & 41.2$\pm$0.7  & 12.81$\pm$0.03  & 12.96$\pm$0.03 & 1190$\pm$80  & 11.45$\pm$0.06 & 11.60$\pm$0.06 &  51$\pm$7  \\
HATLASJ125632.7+233625  & 40.0$\pm$0.6  & 13.11$\pm$0.11  & 13.22$\pm$0.11 & 2140$\pm$470 & 11.76$\pm$0.13 & 11.87$\pm$0.13 &  96$\pm$32 \\
HATLASJ132630.1+334410  & 38.6$\pm$0.6  & 13.21$\pm$0.09  & 13.27$\pm$0.09 & 2440$\pm$430 & 11.59$\pm$0.10 & 11.65$\pm$0.10 &  59$\pm$16 \\
HATLASJ133008.4+245900  & 44.4$\pm$0.8  & 12.66$\pm$0.05  & 12.95$\pm$0.05 & 1150$\pm$120 & 11.52$\pm$0.07 & 11.80$\pm$0.07 &  82$\pm$15 \\
HATLASJ133649.9+291801  & 36.0$\pm$0.7  & 12.93$\pm$0.03  & 12.93$\pm$0.03 & 1090$\pm$80  & 11.78$\pm$0.05 & 11.77$\pm$0.05 &  77$\pm$11 \\
HATLASJ134429.4+303036  & 38.1$\pm$0.4  & 12.89$\pm$0.05  & 12.94$\pm$0.05 & 1140$\pm$130 & 11.73$\pm$0.08 & 11.79$\pm$0.08 &  80$\pm$16 \\
HATLASJ142413.9+022303  & 39.6$\pm$1.0  & 13.20$\pm$0.08  & 13.32$\pm$0.08 & 2740$\pm$450 & 11.91$\pm$0.06 & 12.03$\pm$0.06 & 139$\pm$28 \\
\hline
\end{tabular}
\end{table*}
\begin{table*}
\centering
\caption{
  Same as in Table\,\ref{tab:source_properties_3sigma}, but this time assuming $\mu=\mu_{5\sigma}$ and $A_{\rm dust}=A_{\rm dust, 5\sigma}$. The
  dust temperature is not listed here because it does not depend on the magnification, unless differential magnification is affecting the far-IR 
  to sub-mm photometry.}\label{tab:source_properties_5sigma}
\small
\begin{tabular}{lccccccc}
\hline
IAUname         & $\log{{\rm L}_{\rm FIR}}$   & $\log{{\rm L}_{\rm IR}}$  & SFR     & $\log{\Sigma_{\rm FIR}}$   & $\log{\Sigma_{\rm IR}}$    & $\Sigma_{\rm SFR}$  \\
              & (${\rm L}_{\sun}$)       & (${\rm L}_{\sun}$)     & (${\rm M}_{\sun}{\rm yr}^{-1}$)   & (${\rm L}_{\sun}$\,kpc$^{-2}$) & (${\rm L}_{\sun}$\,kpc$^{-2}$) & (${\rm M}_{\sun}{\rm yr}^{-1}$\,kpc$^{-2}$) \\
\hline
HATLASJ083051.0+013225  & 13.17$\pm$0.08  & 13.46$\pm$0.08 & 3720$\pm$650 & 11.83$\pm$0.06  & 12.11$\pm$0.06  & 167$\pm$35      \\
HATLASJ085358.9+015537  & 12.95$\pm$0.18  & 12.98$\pm$0.18 & 1250$\pm$430 & 11.98$\pm$0.21  & 12.01$\pm$0.21  & 132$\pm$69      \\
HATLASJ090740.0$-$004200& 12.56$\pm$0.08  & 12.81$\pm$0.08 &  840$\pm$150 & 11.55$\pm$0.09  & 11.80$\pm$0.09  &  82$\pm$22      \\
HATLASJ091043.1$-$000321& 12.72$\pm$0.05  & 12.81$\pm$0.05 &  840$\pm$100 & 11.98$\pm$0.15  & 12.07$\pm$0.15  & 152$\pm$48      \\
HATLASJ120127.6$-$014043& 13.12$\pm$0.09  & 13.10$\pm$0.09 & 1650$\pm$320 & 11.87$\pm$0.06  & 11.86$\pm$0.06  &  95$\pm$23      \\
HATLASJ125135.4+261457  & 12.77$\pm$0.04  & 12.92$\pm$0.04 & 1090$\pm$90  & 11.60$\pm$0.07  & 11.75$\pm$0.07  &  72$\pm$12      \\
HATLASJ125632.7+233625  & 13.04$\pm$0.12  & 13.15$\pm$0.12 & 1840$\pm$450 & 12.00$\pm$0.16  & 12.11$\pm$0.16  & 166$\pm$65      \\
HATLASJ132630.1+334410  & 13.20$\pm$0.08  & 13.27$\pm$0.08 & 2410$\pm$400 & 11.73$\pm$0.10  & 11.79$\pm$0.10  &  81$\pm$21      \\
HATLASJ133008.4+245900  & 12.65$\pm$0.05  & 12.94$\pm$0.05 & 1120$\pm$110 & 11.72$\pm$0.12  & 12.01$\pm$0.12  & 132$\pm$35      \\
HATLASJ133649.9+291801  & 12.88$\pm$0.05  & 12.88$\pm$0.05 &  980$\pm$100 & 12.02$\pm$0.06  & 12.01$\pm$0.06  & 134$\pm$23      \\
HATLASJ134429.4+303036  & 12.86$\pm$0.06  & 12.91$\pm$0.06 & 1060$\pm$140 & 11.90$\pm$0.09  & 11.95$\pm$0.09  & 116$\pm$27      \\
HATLASJ142413.9+022303  & 13.26$\pm$0.06  & 13.38$\pm$0.06 & 3130$\pm$400 & 12.30$\pm$0.08  & 12.42$\pm$0.08  & 344$\pm$71      \\
\hline
\end{tabular}
\end{table*}

\subsection{Source magnifications and sizes}\label{subsec:magn_sizes}

Fig.\,\ref{fig:magn_profiles} shows, for each source, how the value of the magnification varies with the size of the region in the SP, as defined 
by the SNR of the pixels and here expressed in terms of $r_{\rm eff}$. The values of $\mu_{3\sigma}$ and $\mu_{5\sigma}$ are shown at the 
corresponding effective radii (light and dark blue squares, respectively), together with the magnification factor estimated by B13 (red dots). 
The latter is placed at a radius $r_{\rm eff}=2\times r_{\rm half}$, where $r_{\rm half}$ is the mean half-light radius of the S{\'e}rsic profile used
by B13 to model the source surface brightness. It is calculated as $r_{\rm half}= a_{\rm s}\sqrt{1-\epsilon}$, with $a_{\rm s}$ and $\epsilon$ 
being the half-light semi-major axis length and ellipticity of the S{\'e}rsic profile provided by B13. B13 computed the magnification factor 
for an elliptical aperture in the SP with semi-major axis length equal to $2\times a_{\rm s}$. It is easy to show that the area of this region is
exactly equal to $\pi \times (2 \times r_{\rm half})^{2}$. \\
In general we find lower values of the magnification factor compared to B13. Discrepancies are to be expected for systems like HATLASJ083051.0+013225
and HATLASJ142413.9+022303, where the best-fitting lens model parameters differ significantly from those of B13. However a similar explanation may be
also applied to systems like HATLASJ085358.9+015537,HATLASJ091043.0$-$000322, and HATLAS134429.4+303034. In HATLAS125135.3+261457,
HATLASJ125632.4+233626 and HATLASJ133008.4+245900 the reconstructed source morphology is indicative of the presence of two partially blended components.
The complexity of the source is not recovered by a single S{\'e}rsic profile and a significant fraction of the source emission lies beyond the region
defined by B13 to compute $\mu$. As a consequence their magnification factor is higher than our estimate. HATLASJ090740.0$-$004200,
HATLASJ132630.1+334410 and HATLASJ133649.9+291800 are the only systems where our findings are quite consistent with those of B13 for both the source
size and the magnification.\\

In Fig.\,\ref{fig:reff_vs_Lir} we show the effective radius of the dust emitting region in DSFGs at $1.5\lsim z \lsim5$ from the literature, as a
function of their infrared luminosity ($L_{\rm IR}$, integrated over the rest-frame wavelength range $8-1000\,\mu$m). Most of these estimates are
obtained from ALMA continuum observations by fitting an elliptical Gaussian model to the source surface brightness. The value of $r_{\rm eff}$ reported
in the figure is the geometric mean of the values of the FWHM of the minor and major axis lengths, unless otherwise specified. We provide below a brief
description of the source samples presented in Fig.\,\ref{fig:reff_vs_Lir}. 

\cite{2015ApJ...807..128S} carried out ALMA follow-up observations at 870$\micron$, with $\sim0.3^{\prime\prime}$ resolution, of 52 DSFGs selected 
from the SCUBA-2 Cosmology Legacy Survey (S2CLS). They provide the median value of the FWHM of the major axis for the sub-sample of 23 DSFGs 
detected at more than 10$\,\sigma$ in the ALMA maps: FWHM$_{\rm major}=2.4\pm0.2\,$kpc. The median infrared luminosity of the same sub-sample is
$L_{\rm IR}=(5.7\pm0.7)\times10^{12}\,L_{\odot}$. These are the values we show in Fig.\,\ref{fig:reff_vs_Lir} (triangular symbols), bearing in mind
that we have no information on the ellipticity of the sources to correct for. Therefore, when comparing with other data sets, the Simpson et al. point
should be considered as an upper limit.

\cite{2015ApJ...812...43B} have presented ALMA 870$\,\mu$m imaging data, at 0.45$^{\prime\prime}$ resolution, of 29 DSFGs from the 
{\it Herschel} Multi-tiered Extragalactic Survey \cite[HerMES;][]{2012MNRAS.424.1614O}. The sample includes both lensed and unlensed objects. 
Lens modelling is carried out assuming an elliptical Gaussian. The un-lensed galaxies are also modelled with an elliptical Gaussian. 
Their results are shown in Fig.\,\ref{fig:reff_vs_Lir} (square symbols), with FWHM$=2\times r_{s}$, where $r_{s}$ is the geometric mean of the 
semi-axes, as reported in their Table 3. The infrared luminosity of the lensed sources have been corrected for the magnification. We only show the 
sources in their sample that are not resolved into multiple components as no redshift and infrared luminosity are available for the individual 
components. This reduces their sample to nine objects: eight strongly lensed and one un-lensed.

\cite{2015ApJ...810..133I} have exploited ALMA 1.1$\,\mu$m continuum observations to measure the size of a sample of 13 AzTEC-selected DSFGs with
$z_{\rm phot}\sim3-6$ and $L_{\rm IR}\sim2-6\times10^{12}\,L_{\odot}$. They fit the data in the $uv$-plane assuming a symmetrical Gaussian.
In Fig.\,\ref{fig:reff_vs_Lir} we show their findings as FWHM$=2\times R_{c,e}$ (hexagon symbols), where $R_{c,e}$ is the value they quote in their
Table\,1 for the half-width at half-maximum of the symmetric Gaussian profile. Their 1.1$\,\mu$m flux densities have been rescaled to 870$\,\mu$m 
by multiplying them by a factor 1.5 (see Oteo et al. 2017). For most of the sources in the Ikarashi et al. sample the redshift is loosely 
constrained, with only lower limits provided. Therefore we only consider here two sources in their sample with an accurate photometric redshift, 
i.e. ASXDF1100.027.1 and ASXDF1100.230.1.

\cite{2016ApJ...833..103H} used high resolution ($0.16^{\prime\prime}$) ALMA 870$\micron$ continuum observations of a sample of 16 DSFGs with
$1\lsim z\lsim 5$ and $L_{\rm IR}\sim4\times10^{12}\,L_{\odot}$ from the LABOCA Extended {\it Chandra} Deep Field South (ECDFS) sub-mm survey 
\citep[LESS;][]{2013MNRAS.432....2K,2013ApJ...768...91H} to investigate their size and morphology. Their results are represented by the diamond
symbols in Fig.\,\ref{fig:reff_vs_Lir}.

\cite{2017arXiv170904191O, 2016ApJ...827...34O} have performed ALMA 870$\,\mu$m continuum observations, at $\sim0.12^{\prime\prime}$ resolution,
of 44 ultrared DSFGs (i.e. with {\it Herschel}/SPIRE colors: $F_{\rm 500\mu m}/F_{\rm 250\mu m}>1.5$ and $F_{\rm 500\mu m}/F_{\rm 350\mu m}>1$).
They confirmed a significant number of lensed galaxies, which we do not consider here because no lens modelling results are available for them yet.
We only consider un-lensed objects for which Oteo et al. provide a photometric or a spectroscopic redshift
\citep{2017arXiv170509660R, 2017MNRAS.472.2028F, 2016ApJ...827...34O}. When a source is resolved into multiple components, each component is fitted
individually and an estimate of the SFR (and, therefore, of ${\rm L}_{\rm IR}$) is provided based on the measured 870$\,\mu$m flux density. The circles in
Fig.\,\ref{fig:reff_vs_Lir} show their findings.

In order to compare with the data from literature we also fit our reconstructed source surface brightness using an elliptical Gaussian model. The derived
values of the FWHM along the major and the minor axis of the ellipse are reported in Table\,\ref{tab:source_properties1} together with their geometric
mean FWHM$_{\rm m}=\sqrt{{\rm FWHM_{maj}}\times{\rm FWHM_{min}}}$. However we warn the reader that the use of a single Gaussian profile to model the
observed surface brightness of DSFGs could bias the inferred sizes because of the clumpy nature of these galaxies, as partially revealed by our SMA data.
In fact, we find that the values of ${\rm FWHM_{\rm m}}$ are systematically lower than those of $r_{\rm eff,5\sigma}$ and $r_{\rm eff,3\sigma}$, as
demonstrated in Fig.\ref{fig:reff_vs_FWHM}.

With this caveat in mind, we show in Fig.\,\ref{fig:reff_vs_Lir} the size of the dust emitting region derived from the Gaussian fit to our reconstructed
source surface brightness. The infrared luminosity, obtained from a fit to the observed spectral energy distribution (see Section\,\ref{subsec:Lir} for
details), has been corrected for lensing by assuming\footnote{Note that, according to Fig.\,\ref{fig:magn_profiles}, the magnification factor does not
change significantly between the scales $r_{\rm eff}=r_{\rm eff,5\sigma}$ and $r_{\rm eff}={\rm FWHM_{\rm m}}$} $\mu=\mu_{5\sigma}$. \\
In Fig.\,\ref{fig:reff_vs_Lir} the data points are coloured according to their redshift. Most of the objects have a photometric redshift estimate;
those with a spectroscopic redshift are highlighted by a black outline. We observe a significant scatter in the distribution of the source sizes,
particularly at the lowest luminosities, with values ranging from $\lsim0.5\,$kpc to $\gsim3\,$kpc. The lack of sources with $r_{\rm eff}\lsim1\,$
kpc at ${\rm L}_{\rm IR}\gsim10^{13}\,{\rm L}_{\sun}$ is possibly a physical effect. In fact, such luminous sources would have extreme values of the SFR surface
brightness, and, therefore, would be quite rare. The absence of $z>3.5$ sources with $r_{\rm eff}\gsim1.5\,$kpc and
${\rm L}_{\rm IR}\lsim3\times10^{12}\,{\rm L}_{\sun}$ is likely due to their lower surface brightness, which makes these objects difficult to resolve in high
resolution imaging data. Based on these considerations it is challenging to draw any conclusion about the dependence of the size on either 
luminosity or redshift.

The sources in our sample have a median effective radius $r_{\rm eff,5\sigma}\sim1.77\,$kpc, rising to $r_{\rm eff,3\sigma}\sim2.46\,$kpc if we 
consider all the pixels in the SP with SNR\,$>3$, while the median FWHM of the Gaussian model is $\sim1.47\,$kpc. These values are consistent with what
observed for other DSFGs at similar, or even higher, redshifts.

\subsection{Star formation rate surface densities}\label{subsec:Lir}

We derive the star formation rate, SFR, of the sources in our sample from the magnification-corrected IR luminosity, ${\rm L}_{\rm IR}$, 
using the \cite{2012ARAA..50..531K} relation:
\begin{equation}
\label{eq:KennicutSFR}
{\rm SFR}\,({\rm M}_{\sun}\,{\rm yr^{-1}}) \sim 1.3\,\times\,10^{-10}\,{\rm L}_{\rm IR}\,({\rm L}_{\sun})
\end{equation}
which assumes a Kroupa initial mass function (IMF). B13 provide an estimate of the total far-infrared (FIR) luminosity, ${\rm L}_{\rm FIR}$
(integrated over the rest-frame wavelength range $40-120\,\mu$m), of the sources in their sample by performing a fit to the measured
{\it Herschel}/SPIRE and SMA photometry using a single temperature, optically thin, modified blackbody spectrum with dust emissivity index
$\beta=1.5$. The normalization of the spectrum and the dust temperature, ${\rm T}_{\rm dust}$, were the only free parameters. We have repeated that 
exercise using the {\it Herschel}/SPIRE photometry from the latest release of the {\it H}-ATLAS catalogues
\citep[][Furlanetto et al. in prep.; as listed in N17, their Table\,4]{2016MNRAS.462.3146V}, including also the {\it Herschel}/PACS photometric 
data points, where available. The fit is performed using a Monte Carlo approach outlined in N17, to account for uncertainities in the photometry 
and in the redshift when the latter is not spectroscopically measured, as in the case of HATLASJ120127.6$-$014043. The observed spectral energy 
distribution (SED) and the best-fitting model are shown in Fig.\,\ref{fig:SEDs}. 

The inferred infrared luminosities and star formation rates are listed in Table\,\ref{tab:source_properties_3sigma} and they have been 
corrected for the effect of lensing by assuming $\mu=\mu_{3\sigma}$ (see Table\,\ref{tab:source_properties1}). To directly compare with 
B13 we also report, in the same table, the magnification-corrected far-IR luminosity. Table\,\ref{tab:source_properties_5sigma} shows the same 
results but corrected assuming $\mu=\mu_{\rm 5\sigma}$. The dust temperature is not listed in that table because it does not depend on lensing, 
unless differential magnification is affecting the far-IR to sub-mm photometry, thus biasing the results of the SED fitting. Unfortunately we 
cannot test if this is the case with the current data.

In both Table\,\ref{tab:source_properties_3sigma} and Table\,\ref{tab:source_properties_5sigma} we report the dust luminosity and star formation
rate surface densities, defined as $\Sigma_{\rm IR}={\rm L}_{\rm IR}/A_{\rm dust}$ and $\Sigma_{\rm SFR}={\rm SFR}/A_{\rm dust}$ respectively. Both are
corrected for the magnification and computed using the value of $A_{\rm dust}$ corresponding to the adopted SNR threshold in the SP.

Fig.\,\ref{fig:SigmaSFR_vs_Lir} shows the SFR surface density of the sources in our sample as a function of their infrared luminosity (squares).
We find median values $\Sigma_{\rm SFR,FWHM}=215\pm114\,M_{\odot}\,$yr$^{-1}$kpc$^{-2}$ (dark magenta line) inside the region of radius
$r={\rm FWHM_{\rm m}}$, and $\Sigma_{\rm SFR,5\sigma}=132\pm69\,M_{\odot}\,$yr$^{-1}$kpc$^{-2}$ (dashed blue line) and
$\Sigma_{\rm SFR,3\sigma}\sim78\pm25\,M_{\odot}\,$yr$^{-1}$kpc$^{-2}$ (dotted cyan line) inside the regions in the source plane with ${\rm SNR\geq5}$
and ${\rm SNR\geq3}$, respectively. The red circles are the findings of B13 for the same sources. We have computed them by taking
the FIR luminosity quoted by B13 and first converting it into ${\rm L}_{\rm IR}$ (by multiplying ${\rm L}_{\rm FIR}$ by a factor 1.9, as reported in B13) and
then into SFR using Eq.\,(\ref{eq:KennicutSFR}). Then we have divided the SFR by the source area calculated as $A_{\rm dust}=\pi r_{\rm half}^{2}$.
Finally we have divided the result by 2. In fact, by definition, the region within the circle of radius $r_{\rm half}$ contributes only half of
the total luminosity (and therefore SFR) of the source. The median value of the SFR surface densities calculated in this way is
$\Sigma_{\rm SFR}\sim219\,M_{\odot}\,$yr$^{-1}$kpc$^{-2}$ (dot-dashed red line) which is similar to our estimate inside the region of radius FWHM,
although the data points of B13 (red circles) display a much larger scatter than ours, and higher than our estimate inside the region defined with
${\rm SNR}$. It is notable that the SFR surface density calculated in this way $-$ although consistent with what done in other works
\citep[e.g.][]{2015ApJ...807..128S,2017ApJ...837..182O} $-$ is not representative of the galaxy as a whole, but only of the central region,
where the emission is likely to be more concentrated. Therefore such an estimate should be taken as an upper limit for the SFR surface brightness
of the whole galaxy. However we cannot exclude that individual star forming regions, resolved in higher resolution imaging data, may show significantly
higher values of $\Sigma_{\rm SFR}$.

We also show, in the same figure, the median SFR surface density of DSFGs from the \cite{2015ApJ...807..128S} sample (green
triangle). They estimated $\Sigma_{\rm SFR}=90\pm30\,M_{\odot}\,$yr$^{-1}$kpc$^{-2}$, assuming a Salpter IMF, which decreases to 
$\Sigma_{\rm SFR}\sim67\,M_{\odot}\,$yr$^{-1}$kpc$^{-2}$ if we assume a Kroupa IMF as in Eq.\,(\ref{eq:KennicutSFR}). Their estimate of the SFR 
surface density is consistent with ours, although their sample has a lower infrared luminosity on average. However their way of calculating
$\Sigma_{\rm SFR}$ is exactly the same we have adopted to draw the B13 data points in Fig.\,\ref{fig:SigmaSFR_vs_Lir} and therefore it is affected
by the same caveat discussed above.

The solid yellow line, at $\sim1000\,M_{\odot}\,$yr$^{-1}$kpc$^{-2}$, marks the theoretical limit for the SFR surface density in a radiation 
pressure supported star forming galaxy \citep{2005ApJ...630..167T,2011EAS....52..275A}. None of our sources are close to that limit, at variance 
with what was found by B13. However we cannot exclude, as mentioned before, that individual star forming regions, not fully resolved by our current data,
may reach the theoretical limit or even exceed it. 

\section{Conclusions}\label{sec:conclusions}

We have reassessed the lens modelling and source reconstruction performed by \cite{2013ApJ...779...25B} on the SMA observations of a sample of 11
lensed galaxies selected from the {\it H}-ATLAS. We have also presented new SMA observations of a further seven candidate lensed galaxies from the
{\it H}-ATLAS sample which allowed us to confirm the lensing in at least one case, which we have included in the lens modelling.

Our lens modelling is based on the Regularized Semilinear Inversion method described in \cite{2003ApJ...590..673W} and \cite{2015MNRAS.452.2940N}, modified
to deal directly with the observed visibilities in the {\it uv} plane. This is a semi-parametric method, meaning that the source surface brightness
counts are retrieved by pixelizing (or tesselating) both the observed image plane and the source plane. This differs from what done in B13, where the
source was assumed to be described by a single S{\'e}rsic profile. In this way, we are able to retrieve the original source morphology, which, for 
these kind of sources, is usually clumpy.

As expected, when the reconstructed source does not display complex morphologies, our results for the lens mass model agree in general with those 
of B13. The only exceptions involve the modelling of multiple lens systems (just two in our sample), where degeneracies between model parameters 
are more likely to occur.

The adopted source reconstruction technique allows us to define a signal-to-noise ratio map in the source plane. We use it to more robustly define the
area of the dust emitting region in the source plane and its corresponding magnification, while in B13, the source extension is an arbitrary factor of
the half-light radius of the adopted S{\'e}rsic profile. We report the size of the reconstructed sources in our sample as the radius of a circle that 
encloses all the source plane pixels with ${\rm SNR}>5$ (or ${\rm SNR}>3$). However, for a more straightforward comparison with results in literature,
we also quote the value of the FWHM obtained from a Gaussian fit to the reconstructed source plane. For almost 50 percent of our sample the estimated
effective radii are larger than $2\times r_{\rm half}$, i.e. the radius of the region chosen by B13 to represent the source physical size when computing
the magnification. As a consequence, we estimate, in general, lower magnification factors than those quoted in B13.

Once corrected for the magnification, our sources still retain very high star formation rates ${\rm SFR}\sim900-3500\,M_{\odot}\,$yr$^{-1}$. With a
median effective radius $r_{\rm eff ,5\sigma}\sim1.77\,$kpc ($r_{\rm eff,3\sigma}\sim2.46$\,kpc) and a median ${\rm FWHM}\sim1.47\,$kpc, our sample has
a median SFR surface density $\Sigma_{\rm SFR,5\sigma}\sim132\,M_{\odot}\,$yr$^{-1}\,$kpc$^{-2}$
($\Sigma_{\rm SFR,3\sigma}\sim78\,M_{\odot}\,$yr$^{-1}\,$kpc$^{-2}$ or $\Sigma_{\rm SFR,FWHM}\sim215\,M_{\odot}\,$yr$^{-1}\,$kpc$^{-2}$ from the Gaussian
fit). This is consistent with what is observed for other DSFGs at similar redshifts, but it is only a $\sim$10 percent of the limit achievable in a
radiation pressure supported starburst galaxy.

\section*{Acknowledgments}

We thank J. Nightingale for useful suggestions about the use of {\sc multinest} in the lens modelling. We thank the anonymous referee for helpful suggestions and useful comments. \\
MN acknowledges financial support from the European Union's Horizon 2020 research and innovation programme under the Marie Sk{\l}odowska-Curie grant
agreement No 707601.\\ 
IO acknowledges support from the European Research Council (ERC) in the form of Advanced Grant, {\sc cosmicism}.\\
M.J.M.~acknowledges the support of the National Science Centre, Poland through the POLONEZ grant 2015/19/P/ST9/04010; this project has received funding
from the European Union's Horizon 2020 research and innovation programme under the Marie Sk{\l}odowska-Curie grant agreement No. 665778.\\
GDZ acknowledges  support from ASI/INAF agreement n.~2014-024-R.1 and from the ASI/Physics Department of the university of Roma--Tor Vergata agreement 
n. 2016-24-H.0\\
The {\it Herschel}-ATLAS is a project with {\it Herschel}, which is an ESA space observatory with science instruments provided by European-led Principal
Investigator consortia and with important participation from NASA. The {\it H}-ATLAS website is http://www.h-atlas.org/. \\
Some of the data presented herein were obtained at the Submillimeter Array, which is a joint project between the Smithsonian Astrophysical Observatory
and the Academia Sinica Institute of Astronomy and Astrophysics and is funded by the Smithsonian Institution and the Academia Sinica.\\


\end{document}